\documentclass[preprint]{aastex}

\newcommand{\etal}{et~al.}
\newcommand{\fwhm}{\sc fwhm}

\newcommand{\kms}{km~s$^{-1}$}
\newcommand{\jasa}{\it Journal of the American Statistical Association}

\newcommand{\zem}{{z_{\mathrm{em}}}}
\newcommand{\aox}{\alpha_{\mathrm{ox}}}

\newcommand{\rew}{W_{\mathrm{r}}}
\newcommand{\CIVdblt}{\ion{C}{4}~$\lambda\lambda 1548, 1550$}
\newcommand{\MgIIdblt}{\ion{Mg}{2}~$\lambda\lambda 2796, 2803$}
\newcommand{\AlIIIdblt}{\ion{Al}{3}~$\lambda\lambda 1855, 1863$}
\newcommand{\NVdblt}{\ion{N}{5}~$\lambda\lambda 1238, 1 242$}
\newcommand{\OVIdblt}{\ion{O}{6}~$\lambda\lambda 1031, 1037$}
\newcommand{\SiIIdblt}{\ion{Si}{2}~$\lambda\lambda1190, 1193$}
\newcommand{\SiIVdblt}{\ion{Si}{4}~$\lambda\lambda1393, 1402$}
\newcommand{\AlI}{\ion{Al}{1}}
\newcommand{\AlII}{\ion{Al}{2}}

\newcommand{\CII}{\ion{C}{2}}
\newcommand{\CIII}{\ion{C}{3}}
\newcommand{\CIV}{\ion{C}{4}}
 
\newcommand{\HI}{\ion{H}{1}}
\newcommand{\Lya}{{Ly}\,$\alpha$}
\newcommand{\Lyb}{{Ly}\,$\beta$}
\newcommand{\Lyg}{{Ly}\,$\gamma$}
\newcommand{\Hb}{{H}\,$\beta$}
\newcommand{\FeII}{\ion{Fe}{2}}
\newcommand{\MgII}{\ion{Mg}{2}}
\newcommand{\NV}{\ion{N}{5}}
\newcommand{\NIII}{\ion{N}{3}}
\newcommand{\OVI}{\ion{O}{6}}
\newcommand{\SiII}{\ion{Si}{3}}
\newcommand{\SiIII}{\ion{Si}{3}}

\tighten
\begin{document}

\received{28 August 2000}
\accepted{5 October 2000}

\slugcomment{accepted for publication in {\it The Astrophysical Journal}}
\shortauthors{Ganguly \etal}
\shorttitle{NALQSO Origins}


\title{\Large\bf On the Origin of Intrinsic Narrow Absorption Lines in {$z\lesssim1$} QSOs}
\pagestyle{empty}

\author{Rajib~Ganguly, Nicholas~A.~Bond, Jane~C.~Charlton \altaffilmark{1},
Michael~Eracleous, W.~N.~Brandt, and Christopher~W.~Churchill}
\medskip
\affil{\normalsize Department of Astronomy and Astrophysics \\
       The Pennsylvania State University,
       University Park, PA 16802 \\
       e-mail: {\tt ganguly, bond, charlton, mce, niel, cwc@astro.psu.edu}}

\altaffiltext{1}{Center for Gravitational Physics and Geometry,
                 The Pennsylvania State University}

\pagestyle{empty}

\begin{abstract}
We present an exhaustive statistical analysis of the associated
($\Delta~v_{\mathrm{abs}}~<~5000$~\kms), high ionization (\CIV, \NV,
\OVI) narrow absorption line (NAL) systems in a sample of 59 QSOs
defined from the {\it Hubble Space Telescope} QSO Absorption Line Key
Project. The goals of the research were twofold: (1) to determine the
frequency of associated NALs at low redshift and in low luminosity
QSOs; and (2) to address the question of what QSO properties either
encourage or inhibit the presence of associated NAL gas. To that end,
we have compiled the QSO rest--frame luminosities at {2500~\AA},
{5~GHz}, and {2~keV}, spectral indices at {2500~\AA} and {5~GHz}, the
{H$\beta$} emission line {\fwhm}, and the radio core fraction at
observed {5~GHz}. In addition, we have measured the {\CIV} emission
line {\fwhm}. We find 17 associated NALs (16 selected by {\CIV} and 1
selected by {\OVI}) toward 15 QSOs, of which {$\sim10$} are
statistically expected to be intrinsic. From a multivariate clustering
analysis, we find that the QSOs group together (in parameter space)
based primarily on radio luminosity, followed (in order of importance)
by radio spectral index, {\CIV} emission line {\fwhm}, and soft X--ray
luminosity. We find that radio--loud QSOs which have compact radio
morphologies, flat radio spectra [$\alpha(\mathrm{5~GHz}) > -0.5$],
{\it and} mediocre {\CIV~\fwhm} ($\lesssim6000$~\kms) do {\it not}
have detectable associated NALs, down to {$\rew($\CIV$)=0.35$~\AA}. We
also find that BALQSOs have an enhanced probability of hosting
detectable NAL gas. In addition, we find that the velocity
distribution of associated NALs are peaked around the {\it emission}
redshifts, rather than the systemic redshifts of the QSOs. Finally, we
find only one strong NAL [$\rew($\CIV$)\gtrsim 1.5$~\AA] in our low
redshift sample. A comparison with previous higher redshift surveys
reveals evolution in the number of strong NAL systems with
redshift. We interpret these results in the context of an
accretion--disk model. We propose that NAL gas hugs the streamlines of
the faster, denser, low latitude wind, which has been associated with
{\it broad} absorption lines. In the framework of this scenario, we
can explain the observational clues as resulting from differences in
orientation and wind properties, the latter presumably associated with
the QSO radio properties.
\end{abstract}



\section{Introduction}
\pagestyle{myheadings}
\markboth{\small~~Ganguly et al. \hfill QSO--Intrinsic Absorbers~~}
         {\small~~Ganguly et al. \hfill QSO--Intrinsic Absorbers~~}

Ultraviolet resonant narrow (i.e., with velocity widths
{$\lesssim500$~\kms}) absorption lines (NALs) observed in the spectra
of QSOs arise from an eclectic assortment of objects ranging from
intervening galaxies to intracluster gas to the intergalactic
medium. Additionally, the well known broad absorption line (BAL)
phenomenon arises from high--velocity dispersion
($\Delta~v\sim10,000$~\kms) gas intrinsic to the QSO. In the last few
years, some high--redshift NALs have been shown to exhibit the
signature of partial coverage, implying that the gas is close to the
central engine \citep{bs97,bhs97,ham97a,gan99,sp00}. Intrinsic NAL gas
has also been identified through variability of the absorption
profiles on relatively short timescales ($\sim$~months) in the host
QSO rest--frame \citep{ham95,bhs97,ham97b,ham97c,ald97}. For clarity,
we introduce the acronym NALQSO as refering to QSOs with truly
intrinsic NALs, distinguishing them from QSOs in which the intervening
gas is entirely responsible from the appearance of NALs, and from
BALQSOs in which the absorbing gas is intrinsic to the QSO, but has a much
larger velocity dispersion.

Through variability studies and through high--resolution spectroscopy,
intrinsic narrow absorption lines show great promise in providing an
understanding of the physical conditions of the environment in which
QSOs reside. However, before asking questions about what can be
learned about NALQSOs through these types of ventures, one must
address where intrinsic NAL gas is located, what the geometry of the
gas is, and why some QSOs have intrinsic NALs while others do
not. That is, what is the physical origin of NALQSOs?

Preceding the revolution in the QSO absorption line field brought on
by high--resolution spectroscopy, progress toward understanding the
origin of NALQSOs was made through the statistical assessment of the
number of intrinsic systems in subsamples of large surveys of
intermediate to high redshift {$1.4 \lesssim \zem \lesssim 4.1$} QSOs
(where the UV resonant doublets, like {\CIVdblt}, are shifted into the
optical). Specifically, the subsamples were restricted to {\it
associated} NALs (those that appear within {$\sim$5000~\kms} of the
QSO emission redshift). The first of these surveys began with a sample
of QSOs from the Third Cambridge Catalogue of Radio Sources
\citep{wwpt,foltz86,and87}. These studies found that there was a
statistical excess of associated systems selected by the {\CIVdblt}
doublet over what would be expected if NALs were randomly distributed
in space. (Strictly speaking, the velocity range was established {\it
after the fact} using the excess.) Moreover, these studies concluded
that high equivalent width systems ({$\gtrsim1.5$~\AA} in the
rest--frame) were preferentially hosted by radio--loud QSOs. Later
studies employing optically--selected samples of QSOs found no such
excess, not even with radio--selected subsamples
\citep{ysb,sbs88}. The primary difference in the radio properties of
the two types of samples was the radio spectral index. The QSOs from
the radio--selected samples primarily had steep radio spectra while
those from the optically--selected, radio--loud subsample had mostly
flat radio spectra. A later survey of {\MgII} absorbers in a sample of
steep radio spectrum, radio--loud QSOs by \cite{ald94} rediscovered
the excess of associated systems.

Also in recent history, there has been mounting evidence that the
``warm absorption'' gas seen through soft X--rays in AGN may be
connected to the intrinsic gas detected through UV absorption (e.g.,
Mathur, Elvis, \& Wilkes 1995\nocite{mew95}). In an HST/FOS archival
survey of Seyfert 1 galaxies, Crenshaw {\etal} (1999\nocite{cren99})
find a one--to--one correspondence between the objects with X--ray
warm absorbers and those with intrinsic UV absorption. Moreover,
Brandt, Laor, \& Wills (2000\nocite{brandt}, hereafter, BLW) find a
striking relation between the rest--frame equivalent width of {\CIV}
absorption and the optical/X--ray two--point spectral index, $\aox$,
in {$\zem\lesssim0.5$} Palomar-Green QSOs. They have argued that this
relation is driven by absorption, with absorbed objects having both
large {\CIV} equivalent widths and large negative values of {$\aox$}
(corresponding to apparently weak soft X--ray emission relative to the
optical emission).

Motivated by these recent developments, we consider the connection
between intrinsic narrow absorption line gas and the properties of
{\it low redshift} QSOs to determine which properties either encourage
or inhibit the detectable presence of intrinsic NALs. As a by--product
of using a low redshift sample, we extend the analysis to lower
luminosity QSOs -- something difficult to do at high redshift -- and
we investigate trends with redshift.  Similar to the Crenshaw {\etal}
(1999\nocite{cren99}) analysis of Seyfert 1 galaxies, we start with
the archived HST/FOS data set provided by the HST QSO Absorption Line
Key Project \citep{kpi,kpvii,kpxiii}. This QSO sample is unbiased
toward the detection of associated absorption and has uniformly good
{$S/N$} spectra. We define the sample in \S2. In
\S3, we present the results of our efforts to detect associated
absorption in these QSOs. In \S4, we discuss the optical, radio, and
X--ray properties compiled for this study. We perform a multivariate
tree--clustering analysis in \S5 and summarize our results in
\S6. Finally, we consider and discuss the constraints the results
place on scenarios for the narrow absorption line gas in \S7.

\section{Sample Definition}

The QSOs used for this study derive from a subsample of QSOs from the
HST QSO Absorption Line Key Project (hereafter, KP).  We restrict
ourselves to KP observations made in the high resolution modes of the
Faint Object Spectrograph (i.e., with either the G130H, G190H, or
G270H grating and with a slit width {$<0.\arcsec26$}). The resolving
power of these modes is {$R\sim1300$} which corresponds to a velocity
resolution of {$\sim 230$~\kms}. We have excluded spectropolarimetric
and lower resolution observations (e.g. with the G160L grating).  Some
observations \citep{kpxiii} were taken before the installation of
COSTAR. However, as a result of the narrow slit width, we expect the
spectral resolution to be the same. In the interest of uniformity, we
further restrict the sample to those QSOs with spectra that cover at
least the range from the broad {\Lya} to the {\CIV} emission lines.
We call this the {$\alpha$} sample. To address the question of false
detections, we also consider another sample, the {$\beta$--sample, in
which we only use QSOs where the {\CIV} emission line is covered (to
look for {\CIV} doublets only).

Following the historical searches for associated systems at high
redshift, we require that the spectra cover the range of apparent
velocities {$\pm$5000~\kms} relative to the emission redshift. We do
not consider in this study the population of high ejection velocity
absorbers which have been shown to be intrinsic to the QSO
\citep{kp96,ham97b,rich99}.  We obtained the entire
sample of fully reduced KP spectra and continuum fits from B. Jannuzi,
S. Kirhakos, and D. Schneider. The details of the reductions are given
in Schneider {\etal}~(1993\nocite{kpii}).  The KP observations are
described in Bahcall {\etal} (1993,1996\nocite{kpi,kpvii}) and Jannuzi
{\etal} (1998\nocite{kpxiii}).

The 60 QSOs used in this sample are listed in Table~\ref{tab:sample}
along with the QSO emission redshifts, {$V$} magnitudes, and the
sample to which they belong. The QSOs all have {$\zem \lesssim
1.2$}. In 59 cases, the spectra also cover transitions from a higher
ionization species other than {\CIV} ({\NV} or {\OVI}). These 59 QSOs
comprise the {$\alpha$}--sample. There are four BALQSOs in the sample
({PG~$0043+039$}, {PG~$1254+047$}, {PG~$1700+518$}, {PG~$2112+059$})
and two mini--BALQSOs ({PG~$1411+422$}, {PG~$2302+029$}).

Since the original KP sample was selected with regard to only the UV
flux and nothing else, namely the presence of associated narrow
absorption, and our selection criteria are not based on QSO
properties, this subsample (the $\alpha$--sample) should be unbiased
toward the detection of associated narrow absorption. In addition,
since the $\alpha$--sample is at low redshift, the QSOs should be more
representative of the general population of QSOs than the
high--redshift samples \citep{wwpt,ysb,foltz86,and87,sbs88} since
lower luminosities can be reached. (We return to this point in \S6.3.)
The majority of objects in the KP sample were selected from the
Palomar--Green (PG) Catalogue ({$17/59$} QSOs), Third and Fourth (3C
and 4C) Cambridge Catalogues ({$12/59$} QSOs), and the Parkes (PKS)
Catalogue ({$18/59$}). The PG Catalogue is optically selected and
contains sources that have a UV excess. The 3C, 4C, and PKS catalogues
are radio selected, with the 3C and 4C catalogues containing objects
with steeper radio spectra than the PKS objects. As a result, there is
a bias toward radio--loud QSOs (35 out of 59 are radio--loud), but
this does not weaken our results. [We note that a QSO is considered
radio--loud if {$L_\nu(\mathrm{5~GHz})>10 L_\nu(\mathrm{2500~\AA})$}
\citep{kel89}.] Additionally, there should be no {\it a priori} bias
in other QSO properties outside of physical relations to the
aforementioned parameters.

\section{Search for Associated Narrow Absorption Systems}

To determine which QSOs in the {$\alpha$}--sample could reliably be
classified as NALQSO candidates, we searched for associated NALs
selected by {\CIV}, {\NV}, and {\OVI} at higher sensitivity than the
{$4.5\sigma$} line lists presented by Jannuzi {\etal}
(1998\nocite{kpxiii}) (i.e. with a lower detection threshold). We
first identified absorption features in the spectra using the
optimized algorithm developed by Churchill {\etal}
(1999,2000\nocite{weaki,mgari}), which is based upon that of Schneider
{\etal} (1993\nocite{kpii}). To identify systems selected by, for
example, the {\CIVdblt} doublet, we culled through a {$3\sigma$}
detection line list and assumed each is the bluer (and often stronger)
transition (i.e., $\lambda1548$). Using the resulting candidate
redshift, we then looked for the redder (and often weaker) transition
(i.e., $\lambda1550$) at a {$1.5\sigma$} threshold and the {\HI~\Lya}
transition ($\lambda1215$) at a {$3\sigma$} threshold. We restricted
the search for systems to the canonical {$10000$~\kms} window centered
on the emission redshift. We avoided confusion with Galactic
absorption by first identifying the {\Lya}, {\Lyb},
{\CII~$\lambda$1334}, {\CII~$\lambda$1036}, {\CIII~$\lambda$977},
{\CIVdblt}, {\NIII~$\lambda$989}, {\NVdblt}, {\OVIdblt}, {\MgIIdblt},
{\AlI~$\lambda$3083}, {\AlII~$\lambda$1670}, {\AlIIIdblt},
{\SiII~$\lambda$1260}, {\SiIIdblt}, {$\lambda1527$},
{\SiIII~$\lambda$1206}, and {\SiIVdblt} lines at {$z=0$}. As discussed
later, we are 95\% complete toward the detection of absorption lines
with {$\rew($\CIV$)\ge0.35$~\AA}.

If at least one of the high ionization doublets (\CIV, \NV, \OVI) and
{\Lya} was detected , we classified the QSO as a candidate NALQSO. In
Table~\ref{tab:nals}, we list the 15 NALQSO candidates, the absorber
velocity relative to the QSO emission redshift and the rest--frame
equivalent widths (or limits) of the {\CIV}, {\NV}, {\OVI}, and {\Lya}
transitions. The limits reported are the most conservative {3$\sigma$}
equivalent width limits in the {$10000$~\kms} velocity window, which
always occurred near the wings of the emission lines. In
Fig.~\ref{fig:systems}, we show velocity--aligned plots of the ions
detected (at a {$3\sigma$} threshold) for the 15 NALQSO candidates.
The transitions shown are: {\HI}$\lambda$1215(\Lya),
$\lambda$1025(\Lyb), $\lambda$972(\Lyg), {\CII}$\lambda$1334,
{\CIII}$\lambda$977, {\CIVdblt}, {\NVdblt}, {\OVIdblt},
{\AlII}$\lambda$1670, {\SiIII}$\lambda$1206, {\SiIVdblt}.

We note that all but two systems, the {$\Delta v_{\mathrm{abs}} =
-1380$~\kms} system toward {PG~$0953+415$}, and the {$\Delta
v_{\mathrm{abs}} = +80$~\kms} system toward {PG~$1411+422$}, were
detected and reported by Jannuzi \etal~(1998\nocite{kpxiii}). The
NALQSO classification is subject to further studies to determine which
systems are due to gas truly intrinsic to the QSO. For the
{$\beta$}--sample, we dismissed the criterion that {\Lya} be detected
(or covered) in the spectra and only require the {$3,1.5\sigma$}
detection of the {\CIVdblt} doublet. In this sample, although there is
only one additional QSO, there are six additional NALQSO candidates
resulting from the looser detection criteria. We view these other
NALQSO candidates as false detections since no other transition
corroborates the existence of the system. That is, there are six
``false--positives'' if the {\Lya} criterion is not used.

On the issue of our sensitivity toward detecting associates NALs, in
Fig.~\ref{fig:ewhisto}, top panel, we plot the cumulative distribution
of both the {\CIV} equivalent width limits for all QSOs (dashed line),
and the {\CIV} equivalent width for the NALQSO candidates. In the
bottom panel, we show a histogram of {\CIV} equivalent widths for the
candidates. The shaded histograms represent the QSOs in the {$\alpha$}
sample, while the unshaded (solid) histograms represent the
{$\beta$}--sample. The histogram shows a drop in the number of
{\CIV}--selected associated NALs below {$\rew\approx0.25$~\AA} instead
of a steady rise as expected for a power law distribution. However,
since our sensitivity toward detecting associated NALs drops below
95\% around {$\rew($\CIV$)\approx0.35$~\AA} (as seen in the cumulative
distribution), we cannot claim a cutoff in the equivalent width
distribution. (Nevertheless, since the equivalent width limit is often
better than reported as a result of the emission line, it may be that
such a cutoff exists.) The distribution of equivalent widths in the
{$\beta$}--sample shows the number of false detections we would have
had, if the presence of {\Lya} absorption not been used as a
criterion.

We estimate how many of the candidate NALQSOs are likely to be real
(i.e., that the NAL gas is truly intrinsic) based on statistics of the
number of {\CIV}--selected systems expected in the redshift path
searched. The KP analysis found 107 {\CIV}--selected systems in a
redshift path of {$\Delta z=44$} at a {$4.5\sigma$} detection
threshold. Thus, the number of {\CIV}--selected systems per unit
redshift expected at this threshold is
{$dN/dz=107/44=2.4$}. Unfortunately, Jannuzi
\etal~(1998\nocite{kpxiii}) do not take into account the changing
equivalent width limit across the spectra. In addition, the number of
{\CIV} systems includes both intervening and intrinsic systems. So,
given the aforementioned caveats, the proper interpretation of this
number is that in a unit redshift path, one would statistically expect
to find 2.4 {\CIV}--selected systems, independent of origin, at a
{$4.5\sigma$} detection threshold. (In spite of this precaution, we
treat this {$dN/dz$} as that for intervening systems.) The
redshift path for our survey given the sample of 59 QSOs and the
{10,000~\kms} window searched in each is {$\Delta z \approx 2$}. Thus,
we expect to detect {$\approx 5$} {\CIV}--selected systems. Our sample
contains 13 {\CIV}--selected systems detected at {$4.5\sigma$}. So
statistically, about eight may be intrinsic.

There are two NALQSO candidates to which the {$dN/dz$}
estimate does not apply. {PG~$0953+415$} has a weak {\CIV}--selected
system not detected at {$4.5\sigma$}, and {PG~$1407+265$} was
{\OVI}--selected with no detected {\CIV}. We have no way of
determining, with this data set, if these NALs are also
intrinsic. Thus, we expect that, out of 15 NALQSO candidates, about
$8-10$ may be real (i.e., that the NALs are due to intrinsic gas).

\section{QSO Properties}

To determine which QSO properties (or combination thereof) are
important toward defining the population of NALQSOs, we first assemble
QSO properties that past studies have deemed relevant. Motivated by
the studies connecting radio properties and X--ray warm absorption to
intrinsic NALs, we focus our attention on three wavebands: radio,
optical/near ultraviolet, and X--ray. The specific properties gathered
from each waveband and the methods for obtaining each property are
described below. For the analysis, we assume a
{$H_o=75~\mathrm{km~s}^{-1}~\mathrm{Mpc}^{-1}, q_o=0.5, \Lambda=0$}
cosmology. Monochromatic luminosity densities were all computed from
the following formula [adapted from Sramek \& Weedman
(1980\nocite{sw80}) and Charlton \& Turner (1987\nocite{ct87})]:
\begin{eqnarray}
\log~L_\nu & = & \log~F_{\nu_o} + \alpha \log \left [ {{\nu} \over {\nu_o}}
\right ] + \log \left [ {{16\pi c} \over {H_o}} \right ]
\nonumber  \\
           &   & + \log \left [ 1-(1+2q_o\zem)^{-1/2} \right ] \\
           &   & - \alpha \log \left [ 1+\zem \right ]. \nonumber
\end{eqnarray}
For spectral indices, we use the convention
{$F_\nu\propto\nu^\alpha$}.

\subsection{Optical/Near Ultraviolet}

In this waveband, the luminosity density and spectral index are
referred to a rest--frame wavelength of {2500~\AA}. We derived the
continuum flux density power law using measurements from the G190H and
G270H gratings as reported by Jannuzi \etal~(1998\nocite{kpxiii}) and
converted the power law into the QSO rest--frame.

The {\CIV~\fwhm} of each QSO was directly measured from the KP spectra
by fitting, at most, two Gaussians to the broad emission line. The
{\Hb~\fwhm} for all but the highest redshift QSOs ($z\gtrsim0.8$) in the
sample were taken from the literature. Since the {\Hb} line is shifted
into the infrared for those QSOs, measurements on the {\fwhm} are not
yet available. In addition, {\Hb~\fwhm} for {PKS~$0122-00$} and
{PKS~$0439-433$} could not be found. In Table~\ref{tab:props}, the
second, third, and fourth columns list the {2500~\AA} flux density,
luminosity density and spectral index. The {\CIV} and {\Hb} {\fwhm}
are listed in columns 11 and 12. The first number in column 15 codes
the {\Hb~\fwhm} reference.

\subsection{Radio}

The radio properties used in the analysis were the luminosity density
and spectral index at rest--frame {5~GHz}, and the core fraction,
{$R_{\mathrm{c}}$}, defined as the ratio of the core flux density to the
total flux density, measured at an observed frequency of {5~GHz}. The
core fractions were taken from the literature when
available. Otherwise, we adopted a core fraction based on the {5~GHz}
spectral index (1 for flat radio spectrum QSOs; 0 for steep radio
spectrum QSOs).

The radio continuum power law was derived from the flux densities at
observer--frame {1.4~GHz} and {5~GHz}. When that information was
unavailable, we substituted the {2.7~GHz} flux density for the
{1.4~GHz} flux density and/or the {4.85~GHz} flux density for the
{5~GHz} flux density as needed. When there was insufficient information
to compute a spectral index, we adopted a value of {$\alpha=-0.5$} and
no value for the core fraction. In columns five, six and seven of
Table~\ref{tab:props}, we list the rest--frame {5~GHz} flux density,
luminosity density, and the spectral index. The second and third
numbers in column 15 encode the references for the flux densities used
to compute the power law, while the fourth number provides the
reference for the core fraction.

\subsection{Soft X--ray}

From the X--ray regime, we use the luminosity density at {2~keV}. In
most cases, we calculated this quantity from the ROSAT PSPC count rate
using a spectral model expected from ROSAT PSPC studies of other
QSOs. Specifically, we adopted a power law model with Galactic
absorption \citep{dl90} from a solar metallicity ISM and computed the
power law energy index using the {$\alpha$--\Hb~\fwhm} relation from
BLW:
\begin{equation}
\alpha = -6.122 + 1.277 \log \left ( \mathrm{H}\,\beta~\mathrm{FWHM} \right ).
\end{equation}
In cases where the {\Hb~\fwhm} could not be determined, we adopted a
mean value depending on the radio--loudness of the QSO,
{$\alpha=-1.12$} for radio--loud QSOs and {$\alpha=-1.69$} otherwise
\citep{laor97}.

Since the power law energy index was only required as a step in
determining the {2~keV} luminosity density and had not been directly
measured, we do not include it in the subsequent analyses. For
comparison with the Palomar--Green sample from BLW, we computed the
{2500~\AA}--{2~keV} spectral index, {$\aox$} defined as:
\begin{equation}
\aox=-0.384 \left [ \log L_\nu(\mathrm{2500~\mbox{\AA}})-\log
L_\nu(\mathrm{2~keV}) \right ].
\end{equation}
BLW actually compute the {3000~\AA}--{2~keV} spectral index.  The
correction, in the range {$0.004 \leq \Delta \aox \leq 0.07$}, is too
small to merit implementation given the uncertainties in the data. In
Table~\ref{tab:props}, we list the rest--frame {2~keV} flux density
(column 8), {2~keV} luminosity density (column 9), {2500~\AA}--{2~keV}
spectral index (column 10), and soft X--ray count rate reference
(column 15, fifth number).

\subsection{Bivariate Analysis}
In Fig.~\ref{fig:distrib}, we show the distributions of the following
eight measured properties for the QSOs in the {$\alpha$}--sample: (1)
{$\log L_\nu($2500~\AA$)$}; (2) {$\log L_\nu(\mathrm{5~GHz})$}; (3)
{$\log L_\nu(\mathrm{2~keV})$}; (4) {$\alpha($2500~\AA$)$}; (5)
{$\alpha(\mathrm{5~GHz})$}; (6) {\Hb~\fwhm}; (7) {\CIV~\fwhm}; (8)
{$R_{\mathrm{c}}$}. In Fig.~\ref{fig:scatter}, we present bivariate
scatter plots of all combinations these properties. Candidate NALQSOs
are shown as filled symbols, while all other QSOs are depicted as open
symbols. The symbol shapes are discussed in \S5.  In
Fig.~\ref{fig:ewmosaic}, we show, for the associated NALs, the
rest--frame {\CIV} absorption equivalent width against the eight
aforementioned host QSO properties. In Table~\ref{tab:spear}, we show
the Spearman correlation probabilities for each pair of parameters,
including {$\rew($\CIV$)$}. With the exception of
{$L_\nu(\mathrm{2~keV})$} (as discussed by BLW), there are no clear
correlations that show whether or not a QSO is likely to have a
intrinsic NAL. From these plots, there is also no clear region of
parameter space that separates NALQSOs from QSOs without intrinsic
NALs.

\section{Multivariate Analysis}

To explore the multivariate properties of the QSOs, we performed a
hierarchical tree--clustering analysis on the entire sample of 59 QSOs
without regard to the presence of an associated NAL. The purpose of
this type analysis is to describe how objects group together in
parameter space. Unlike a principal component analysis,
tree--clustering does not look for underlying trends or physical
parameters. Instead, the analysis attempts to separate the objects
into different populations. Since tree-clustering is not yet a
standard analysis tool in the field, we describe the technique in the
next two subsections. In the third subsection, we present the number
of groups that the QSOs occupy and provide the physical
characterizations of each group. In addition, we consider the number
of NALQSO candidates in each group to determine which QSO properties
favor or hinder the detectable presence of intrinsic NAL gas.

\subsection{Description of Basic Tree--Clustering}

An example of how a hierarchical clustering analysis works
follows. Assuming {$N$} elements (e.g., QSOs) that are to be
clustered, form an {$N \times N$} matrix composed of the distances (in
parameter space) between each pair of objects. (This matrix is
symmetric and zero along the diagonal.) Cull though the entries and
find the two elements that are closest together (i.e. have the
smallest distance). Record this distance. Replace the two rows and
columns representing distances from the two elements with a single row
and column representing the distances from the union to the remaining
{$N-2$} elements. Repeat until all elements form one group
\citep{jw82}; i.e., until the matrix has a dimension of one. The
record of the unions and their ``linkage'' distances, called the
amalgamation schedule, is the result of the procedure. There are three
issues in executing this type of analysis.  First, what type of
coordinates does one use to represent the positions of objects in the
multidimensional space (e.g., measured, normalized, standardized
values)? Second, how does one define the distance between two points
(i.e., what is the metric of the space)?  Finally, how does one
amalgamate points onto a group (i.e., what is the ``position'' of a
group)?

The linkage distance between two groups can be expressed recursively.
(This is computationally desirable.) If group {$k$} results from the
union of groups {$i$} and {$j$}, then the distance, {$d(h,k)$},
between group {$h$} and group {$k$} is:
\begin{equation}
d(h,k) = {{n(h,i) d(h,i) + n(h,j) d(h,j) - n(i,j) d(i,j)} \over
{n(h,k)}}.
\end{equation}
\citep{zup82} where {$n(x,y)$} is the sum of the number of objects in
groups {$x$} and {$y$}. In this computation, the distance between two
points in parameter space is computed assuming a Euclidean metric.

An interesting boundary occurs at the iteration where the linkage
distance fractionally changes the most. Before this iteration,
groupings of objects are not statistically significant. After this
iteration, the linkage distance changes dramatically signifying that
truly different populations are being represented by a single
group. The important parameters that separate the groups can be
inferred by simply comparing the (one-dimensional) distributions of
each parameter (e.g., with the Kolmogorov--Smirnov test).

\subsection{Ward's Method of Amalgamation}

Ward's method for amalgamation of points takes into account the
distribution of points in a candidate group \citep{ward63}. In this
method, the decision of how elements are grouped together (i.e., the
amalgamation schedule) is based on an objective function, which
describes how much information is lost by grouping. In so doing, the
need to define distances between groups of objects is completely
superseded. The objective function quantifies the variation of
parameters when elements are grouped together. The ideal value of the
objective function is zero. (An example of an objective function is
the standard deviation.)

The basic idea of the method is to determine the optimal grouping of
objects into a predetermined number of groups that minimizes the sum
of the objective function values for each group. This is done by
placing all {$N$} elements in their own group (i.e., {$N$} groups of
one element each) in the first iteration. (The total objective
function is identically zero in this case.) Next, one proceeds to
{$N-1$} groups, which results in {$N-2$} groups with one elements and
one group with two elements. The third iteration, with {$N-2$} groups,
can either lead to {$N-3$} single--element groups and one
triple--element group, or {$N-4$} single--element groups and two
double--element groups. The process continues, decreasing the number
of groups and recording the group memberships until all elements are
in one group.

\subsection{Parameter Space Grouping of QSOs}

Our analysis of the 59 QSOs used standardized coordinates, a Euclidean
metric for distance computation, and Ward's method for
amalgamation. [A standardized coordinate is just the number of
standard deviations away from mean of the distribution of the
coordinate. For example, the standardized luminosity coordinate is
given by: {$(L-\bar{L})/\sigma_L$}. More generally, this is referred
to as an N(0,1) standardization since the mean value is mapped to zero
and a standard deviation is mapped to unity. Standardized coordinates
are generally used: (1) to compare parameters that do not have the
same, or similar, units; and (2) when the data are ``measured on
scales with widely differing ranges'' \citep{jw82}.] Limits on
parameters were treated as values and missing data were substituted by
the mean (i.e., zero in our coordinate standardization). In all, there
were 24 limits and 32 missing values in the set of 472
coordinates. The objective function was defined as the sum of the
variances over each dimension and group
\citep{ward63}.

In Fig.~\ref{fig:cluster}, we show the amalgamation schedule for the
59 QSOs in the {$\alpha$}--sample. The parameters used to define the
multidimensional space are: (1) {$L_\nu($2500~\AA$)$}, (2)
{$L_\nu(\mathrm{5~GHz})$}, (3) {$L_\nu(\mathrm{2~keV})$}, (4)
{$\alpha($2500~\AA$)$}, (5) {$\alpha(\mathrm{5~GHz})$}, (6)
{\Hb~\fwhm}, (7) {\CIV~\fwhm}, and (8) {$R_{\mathrm{c}}$}. The QSOs
appear to separate into five groups with linkage distances (or
characteristic group sizes) {$<7.8$}. In Fig.~\ref{fig:scree}, we show
a scree plot, which shows how the linkage distance changes with
iteration, as well as the fractional changes. The scree plot reveals
that, after the 54th iteration, the union of groups significantly
changes the objective function. That is, truly distinct populations of
QSOs are being juxtaposed. In Table~\ref{tab:props}, we list the group
number for each QSO in column 14.

In Table~\ref{tab:clusprops}, we list the Kolmogorov--Smirnov
logarithmic probabilities (for each parameter) that the two groups of
QSOs listed in each column are drawn from the same parent
distribution. Working from the top of the tree
(Fig.~\ref{fig:cluster}), downward, the first division of the QSOs is
most strongly related to the bimodality of the radio luminosity
function, with groups one and two containing generally radio--quiet
QSOs ({$17/18$} QSOs have {$\log L_\nu(\mathrm{5~GHz}) < 33$}) and
groups three, four, and five containing radio--loud QSOs ({$33/41$}
QSOs have {$\log L_\nu(\mathrm{5~GHz}) > 33$}).  Groups three and four
are separated from group five by both the radio spectral index and
radio core fraction. All 31 QSOs in groups three and four have
{$\alpha(\mathrm{5~GHz}) \gtrsim -0.7$} while {$8/10$} QSOs in group
five have {$\alpha(\mathrm{5~GHz}) \lesssim -0.7$}.  Groups three and
four are distinguished by the {\CIV} emission line {\fwhm}. Group
three QSOs have {\CIV~\fwhm $< 6000$~\kms}, while {$9/13$} group four
QSOs have {\CIV~\fwhm $> 6000$~\kms}.  The division between groups one
and two is apparently the soft X--ray luminosity. Group one has
{$11/12$} QSOs with {$\log L_\nu(\mathrm{2~keV}) > 26$} while all six
QSOs in group two have {$\log L_\nu(\mathrm{2~keV}) < 26$}.

In Fig.~\ref{fig:kmean}, we show a plot of the mean properties of the
five groups for each parameter. In Fig.~\ref{fig:scatter}, the five
different symbol shapes denote the five groups, corresponding to the
symbols used in Fig.~\ref{fig:kmean}. In Table~\ref{tab:cluster}, we
list for the five groups, the number of QSOs in the group (column 2),
the number of candidate NALQSOs (column 3), the fraction of QSOs that
are candidate NALQSOs (column 4) with formal {$1\sigma$} error bars
(based on binomial statistics), the number of NALs (regardless of
origin) expected in the group (column 5), and the probabilities of
finding that many intervening NALs in the search window (column 6) and
of finding that many associated NALs (column 7) if all QSOs were
equally likely to have intervening or associated NALs, respectively.

Of special interest is group three which has no NALQSO
candidates. This group consists of 18 QSOs, so it is very unlikely (a
0.005 probability), and therefore very significant, that no associated
NALs are detected.  The group is distinguished (in order of division)
by its high radio luminosity density, flat radio spectrum, high radio
core fraction, and mediocre {\CIV~\fwhm} ($<6000$~\kms). All of the
other groups have QSOs with associated NALs.

We also note here that group two, which is radio--quiet and soft
X--ray weak, has a high incidence of both narrow and broad absorption
lines. Out of six QSOs in this group, four are NALQSO candidates
({PG~$0043+039$}, {PG~$1411+422$}, {PG~$1700+518$}, {PKS~$2251+11$}),
three are BALQSOs ({PG~$0043+039$}, {PG~$1700+518$}, {PG~$2112+059$}),
and one is a mini--BALQSO ({PG~$1411+442$}). Moreover, three of the
four BALQSOs in this sample (the above and PG~$1254+047$) are also
NALQSO candidates.  Since the probability of randomly drawing four
NALQSO candidates from a sample of six QSOs is 0.035, and the
probability of randomly drawing three NALQSO candidates out of four
BALQSOs is 0.049, it appears that there is an enhanced probability of
finding associated NAL gas when BAL gas is also present.

A clustering analysis which includes the QSO emission redshift does
not change the amalgamation schedule. This points toward one of two
implications. First, there is no apparent evolution of the QSOs {\it
within} the sample. Second, the grouping of QSOs arises as a result of
evolution.

A Kolmogorov--Smirnov test on the redshift distributions show that the
radio--quiet groups (one and two) are statistically different than the
radio--loud groups (three, four, and five). The radio--loud groups are
at {$\bar{z}_{\mathrm{em,345}}=0.66\pm0.05$} while the radio--quiet groups
are at {$\bar{z}_{\mathrm{em,12}}=0.34\pm0.02$}. This can be understood as
a Malmquist bias in the {5~GHz} luminosity density -- that higher
redshift samples contain higher luminosity objects than lower redshift
samples.

In addition, the Kolmogorov--Smirnov test shows that the redshift
distribution of group three is different from group four
($\bar{z}_{\mathrm{em,3}}=0.6\pm0.2$;
$\bar{z}_{\mathrm{em,4}}=0.9\pm0.2$). This is also the result of a
Malmquist bias. Group four has a higher {2500~\AA} luminosity density
[$\bar{L}_{\nu}($2500~\AA$)_{4}=31.2\pm0.3$] than group three
[$\bar{L}_{\nu}($2500~\AA$)_{3}=30.7\pm0.3$]. There are no other
statistically significant redshift--based separations in the groups.

\section{Results}

\subsection{Properties of NALQSOs}

Based on a rough redshift path density ($dN/dz$) estimate (see \S3),
we expect that five of the 15 associated systems detected in the
sample are intervening systems along the line of sight. While it is
not known which are truly intrinsic, we warn that this would dilute
trends in this statistical study.

A close examination of the {$\alpha(\mathrm{5~GHz})~\mathrm{vs.}~\log
L_\nu($2500~\AA$)$} plot (first column, fourth row in Fig. 3a) reveals
that associated NALs appear in QSOs across the range of both
properties. The fraction of QSOs that host associated NALs is higher
for steep radio spectrum QSOs ({$6/16 = 37\pm12\%$}) than for flat
radio spectrum QSOs ({$4/29 = 14\pm6\%$}). (We note here that five
NALQSO candidates do not have measured radio spectral indices.) Using
the {$dN/dz$} estimate, the probability that all four flat radio
spectrum NALQSO candidates are intervening is only 0.004. It is,
therefore, unlikely that only steep radio spectrum QSOs host intrinsic
NALs. In addition, there is no apparent dependence on optical
luminosity. A Kolmogorov--Smirnov test does not show a significant
difference between the two optical luminosity distributions.

However, an analysis that does not fully take into account the
multivariate properties of QSOs can yield erroneous results. This is
most dramatically illustrated by an analysis of the radio core
fraction of the QSOs. In Fig.~\ref{fig:rchisto}, we show the radio
core fraction distributions of the NALQSO candidates (shaded) and the
entire sample (unshaded). Under a Kolmogorov--Smirnov test, the two
distribution do not significantly differ. Furthermore, the fraction of
NALQSO candidates does not statistically change between high ({$6/18$}
for QSOs with {$R_c\geq0.58$}) and low ({$5/19$} for QSOs with
{$R_c<0.58$}) core fraction. Since the radio core fraction is a
generally accepted indicator of inclination \citep{ob82,wb86}, it
appears that NALQSOs have no preferred orientation based upon this
simple bivariate analysis.

A multivariate clustering analysis shows that the orientation result
is oversimplified. Orientation is important, but it is not the only
observational parameter related to the likelihood that an associated
NAL is present in the spectrum of a QSO. The QSOs in this sample
divide into five distinct groups in the space defined by the radio,
optical/UV, and X--ray continuum and broad emission line
properties. One group of 18 QSOs is completely devoid of associated
narrow absorption, {\it intrinsic or otherwise}. Statistically, one
would expect {$1-2$} intervening NALs in the redshift path
searched. If the appearance of associated NALs were independent of the
host QSO properties, the probability of randomly drawing 18 QSOs
without associated NALs is small ($0.005$). Moreover, since this is
the largest of the five groups, the bias of the sample only
strengthens the significance of no associated NALs. (That is, the size
of this group indicates that we are biased toward QSOs that exhibit
the properties that typify the group. If the properties of the group
allowed for the detectable presence of associated NALs, we would have
seen them.)

The QSOs in this group are distinguished by a combination of
properties which include high radio luminosity, flat radio spectral
index (high radio core fraction), and mediocre {\CIV~\fwhm}
($<6000$~\kms). It is this {\it combination} of properties, and not
any one property (or underlying parameter as in a principal component
analysis), that is apparently linked to the lack of detected
associated NALs in the group. The result, of course, is subject to
follow--up observations of the 18 QSOs to search for even weaker
systems, but holds for {$\rew($\CIV$)>0.35$~\AA} associated NALs.

\subsection{Associated NAL Ejection Velocities}

In Fig.~\ref{fig:vejhisto}, we show a histogram of the absorber
velocities relative to the QSO--emission redshift (determined from the
UV emission lines). The unshaded histogram represents all the
associated NALs while the shaded histogram represents the
{$\rew($\CIV$)>0.35$~\AA} associated NALs (for which we are 95\%
complete). In the absence of knowing which systems are intrinsic, it
is interesting to note that the number of systems with
{$|v_{\mathrm{ej}}| > 1300$~\kms} is consistent with the number of
expected intervening systems. One is tempted to speculate that the
velocity distribution of possibly intrinsic systems is, then,
consistent with the velocity dispersions of rich galaxy clusters (of
which the NALQSO candidates may be a part). If the dispersion were due
to other galaxies in the cluster, however, one would expect the
velocity distribution to peak around the {\it systemic} redshift of
the QSO, not the UV emission redshift. As others have shown [e.g.,
Espey (1993\nocite{esp93}), and references therein], the redshift
derived from UV emission lines is usually {\it blueshifted} relative
to the systemic redshift by over {1200~\kms}. If the
{$|v_{\mathrm{ej}}| < 1300$~\kms} systems are shown to be truly due to
QSO--intrinsic gas, this points toward a striking connection between
the velocity of the emission line peak and the ``ejection'' velocity
of intrinsic NAL gas.

\subsection{Evolution of Strong Associated NALs}

In addition to the lack of associated NALs in group three, this entire
low redshift sample is devoid of the ``strong,'' associated systems
seen in higher redshift samples.  [A strong system is one in which
{$\rew($\CIV$) \ge 1.5$~\AA}.]  In the sample of 24 intermediate
redshift 3C QSOs, Foltz \etal~(1986\nocite{foltz86}) (hereafter
FWPSMC) found 8 strong NALQSO candidates. In our sample, only one
NALQSO candidate ({3C~351} with {$\rew($\CIV$)=2.7$~\AA}) qualifies as
a strong system.

There are two possible reasons for this disparity. The first is that
strong systems have evolved out of existence from intermediate to low
redshift. The other is that strong systems are detected in higher
luminosity QSOs. That is, since it is harder to detect low luminosity
QSOs at intermediate redshift than at low redshift, the intermediate
redshift samples are filled with higher luminosity QSOs.

In Fig.~\ref{fig:foltzcomp}, we show distributions of
{$L_\nu($2500~\AA$)$}, {$L_\nu(\mathrm{5~GHz})$}, and {$\log
R^*$} (The radio--loudness parameter, $R^*$, is defined as the
radio--to--optical luminosity density ratio,
$L_\nu(\mathrm{5~GHz})/L_\nu($2500~\AA$)$.)  of our sample
(unshaded) and the FWPSMC sample (shaded). Our sample is clearly more
representative of lower luminosity QSOs. This does not explain the
disparity, however. If we restrict the comparison to a subsample of
overlapping luminosity densities [say {$L_\nu($2500~\AA$)>30.5$}
and {$L_\nu(\mathrm{5~GHz})>32.5$}], the disparity is {\it more}
striking. There are no strong systems at low redshift in the same
luminosity range as the FWPSMC sample.

Furthermore, we can also further restrict the subsample to include
only steep radio spectrum sources -- unlike the comparisons made by
Sargent \etal~(1988\nocite{sbs88}). The QSOs from group five of the
clustering analysis satisfy both the luminosity criteria and the steep
radio spectrum criterion (compare Fig.~\ref{fig:kmean} with
Fig.~\ref{fig:foltzcomp}). This group has 13 QSOs of only one is a
strong NALQSO candidate (3C~351).  If there were no evolution of
strong systems, we would expect to find {$\sim4$} strong NALQSOs in
this group. If the probability of a QSO in this group hosting a strong
associated NAL is accurately given by the FWPSMC statistics ($8/24$),
then there is only a 0.03 probability of drawing 13 QSOs with only one
hosting a strong associated NAL. Thus, it is likely that strong
associated NALs have largely evolved away. This may have implications
about the fueling history of QSO central engines, as we discuss
further in
\S7.2.

\subsection{The UV/X--ray connection}

Of the four BALQSOs and two mini--BALQSOs in this sample, four
({PG~$0043+039$}, {PG~$1700+518$}, {PG~$1411+442$}, {PG~$2302+029$})
show evidence of narrow absorption line gas. (There is a 0.0037
probability that all are intervening NALs.) Three of the four BALQSOs
appear in the second group (see Fig.~\ref{fig:cluster},
Table~\ref{tab:cluster}, and Fig.~\ref{fig:kmean}).  ({PG~$1254+047$}
is the exception. The unrestrictive limits, which were treated as
values by the clustering algorithm, allowed this QSO to appear in
group four.) This is not unexpected since the broad absorption line
gas tends to wipe out X--ray continuum photons. The optical/X--ray
spectral index for group two is {$\aox=-1.74$} ($\sigma_{\aox}=0.33$).
In Fig.~\ref{fig:niel}, we reproduce the
{$\rew$(\CIV)}--{$\aox$} plot from BLW and superimpose the
QSOs in our sample. When one uses the BAL equivalent width, our data
are largely consistent with BLW, who found an anti--correlation
between the {\CIV} absorption equivalent width and {$\aox$}.

There are three notable exceptions. The {$\aox$} we derive for
{PKS~$2251+11$} is flatter (resulting from the smaller optical flux
derived from the Key Project calibrations), whereas the equivalent
width is the same. This change makes the QSO more consistent with the
general anti--correlation. In reality, since the {\CIV} emission line
of {PKS~$2251+11$} appears asymmetric, there may be a BAL-like outflow
resulting in a higher {\CIV} equivalent width.

{NGC~$2841$~UB3} and {3C~$232$} both fall below the general
{$\rew($\CIV$)$}--{$\aox$} anti--correlation. {3C~$232$} is
actually a QSO--galaxy pair. If the galaxy is in front of the QSO, it
may remove some of the X--ray flux giving an anomalously steep
spectral index. In this case, it also seems plausible that at least
some of the {\CIV} equivalent width derived for the QSO should be
attributed to the galaxy.

{NGC~$2841$~UB3} did not have any detected absorption in the
{$\pm$5000~\kms} window, but the soft X--rays are apparently absorbed
($\aox\sim-1.9$). A cursory look at the Jannuzi {\etal}
(1998\nocite{kpxiii}) line list reveals a possible {\CIV} absorption
system at {$z=0.5116$} (the weaker transition is blended at this
resolution with Galactic {\FeII}) with a rest--frame equivalent width
of {$\rew=0.82$~\AA}. If this system is intrinsic (with a
velocity of {$\Delta v_{\mathrm{abs}}=-8100$~\kms}, outside our search
window), it might explain the weakness of the soft X--rays.

\section{Discussion}
\subsection{Model Requirements from the Properties of NALQSOs
and the Disk-Wind Model}

From the preceding sections, there are four major attributes that any
model attempting to explain NALQSOs, or QSOs in general, must
exhibit. First, and foremost, QSOs that have properties placing them
in group three (i.e., radio--loud, flat radio spectrum, radio core
dominated, mediocre {\CIV~\fwhm}) should be noticeably devoid of
intrinsic NALs with {$\rew($\CIV$)\gtrsim0.35$~\AA}. This can be
accomplished either through the absence of intrinsic NAL gas or
through an observational selection effect (e.g., orientation or small
equivalent width).  Second, the velocity field of intrinsic NAL gas
should be such that, when projected onto the line of sight, intrinsic
NALs may appear either blueward or redward of the UV emission line
peak. But the distribution of line of sight velocities should be
roughly centered about the QSO emission redshift. Third, intrinsic NAL
gas should evolve such that strong [$\rew($\CIV$)>1.5$~\AA] systems
are conspicuously rare at low redshift ($\zem\lesssim1$). Finally,
there should be a connection between soft X--ray weak QSOs and the
strength of UV absorption and an enhanced probability of detecting
intrinsic NAL gas in BALQSOs.

To explain the above properties, we consider the disk--wind model of
QSOs. We begin with a brief summary of the disk--wind model that has
been used to explain the existence of BALQSOs \citep[hereafter
MCGV]{mur95}. In this model, a wind is launched from the entire
surface of the accretion disk. The dynamics of the wind are
effectively determined by radiation pressure. Local radiation pressure
elevates the atmosphere of the disk over a range of radii, while UV
photons originating from the innermost parts of the disk provide the
radiation pressure needed to accelerate this gas radially outwards
until it reaches a terminal speed.  As the wind expands from the
center it conserves angular momentum and therefore continues to rotate
out to a large distance, although at an ever decreasing rate.  The
inner part of the wind is the source of the broad UV emission
lines. When viewed from large inclination angles, absorption by this
high-velocity dispersion wind produces deep, broad troughs, resulting
in a BALQSO. The detailed dynamics have been worked out analytically
by MCGV and simulated numerically by Proga, Stone, \& Kallman
(2000\nocite{psk00}; hereafter, PSK).  An example of the resulting
geometry is shown in the cartoon of Fig.~\ref{fig:diskwind}. With
these theoretical ideas in mind, we consider how the observed
intrinsic NAL properties can be explained in the context of wind
models.

\subsection{One Possible Explanation}

If we postulate that the gas producing intrinsic NALs exists in all
QSOs (in the spirit of perpetuating unification schemes), a disk--wind
scenario, much like the one described above, offers a framework within
which the properties of intrinsic NALs can be understood. In
Fig.~\ref{fig:diskwind}, we show a cartoon of the disk--wind model. In
this scheme, intrinsic NAL gas exists in a region hugging the wind
streamlines at relatively large distances from the continuum
source. The simulations of PSK show that the region in the polar
direction is also occupied by gas launched from the disk, but this gas
is highly ionized and hence not an effective absorber. Furthermore, we
hypothesize that there is a difference in the properties of the wind
(density and mass loss rate) between radio--loud and radio--quiet
QSOs. In particular, we propose that, in comparison to radio--loud
QSOs, radio--quiet QSOs have a higher mass loss rate (relative to the
mass accretion rate), a higher wind velocity, and therefore, a higher
density of matter in their winds.

Our proposal is based on the observation that the UV--to--X--ray
spectra of radio--quiet QSOs are steeper than those of their
radio--loud counterparts, meaning that radio--quiet QSOs have stronger
UV bumps than radio--loud ones (see, for example, Green {\etal}
1995\nocite{green95}).  The relative UV--to--X--ray luminosity of a
QSO is a central parameter in the models of MCGV since it determines
how effectively the wind can be accelerated by UV photon pressure and
how vulnerable it is to over--ionization by X--rays.  The reason for
this difference in spectral energy distributions is not important for
the purposes of our argument. However, we are tempted to speculate
that in radio--quiet QSOs, the accretion rate onto the central black
hole is much closer to the Eddington limit than in radio--loud QSOs
(in other words, the ratio $\dot M/\dot M_{\mathrm{Edd}}$ is closer to
unity in the former case and considerably lower than unity in the
latter).

In the context of the picture described above, we may interpret the
five groups from our clustering analysis as the result of a different
radio luminosity (radio--quiet vs. radio--loud) and a different
viewing angle of the central engine by the observer. We illustrate
this interpretation in Fig.~\ref{fig:diskwind} by using arrows to
indicate the direction of the observer corresponding to each group.
An observer at a given orientation is never guaranteed to observe a
NALQSO because the intrinsic NAL gas may be patchy and not cover all
lines of sight in the same direction uniformly.  We elaborate on the
proposed scheme below, starting with groups three, four, and five,
which contain radio--loud QSOs for which the radio morphology can
serve as an orientation indicator, then continuing with the
radio--quiet QSOs.

\begin{itemize}

\item
QSOs in group five are double--lobed radio sources, which suggests
that the viewing angle from the axis of the accretion disk is
large. The observer's line of sight is likely to pass through parcels
of intrinsic NAL gas, with the result that intrinsic NALs appear in the
spectrum. Although at this inclination the observer's line of sight is
likely to pass through the faster, denser parts of the wind, the
density of the fast wind in radio--loud QSOs is low enough, as we have
postulated above, that BALs do not appear. This contention is
supported by the fact that QSOs in this group do not show the
signature of heavy absorption in their spectral energy distributions,
unlike QSOs in group two, which we discuss later on.

\item
Group three represents the opposite extreme in inclination angles. The
compact radio morphology and flat radio spectra of objects in this
group, suggest that the observer's line of sight is very close to the
axis of the radio jet. As a result, the likelihood of photons from the
UV continuum source or broad--emission line region to pass through the
NAL region on their way to the observer is extremely small. Hence, this
group includes no NALQSO candidates.

\item
Group four comprises a mixture of lobe--dominated and core--dominated
radio sources. Thus, we interpret this as a group of QSOs viewed at
intermediate orientations between pole on and edge on with a finite
probability that the line of sight intercepts NAL gas. It is
noteworthy that this group contains the objects with the broadest
{\CIV} emission lines in our entire collection. If we adopt the above
prescription for the dynamics of the wind, then the large width of the
{\CIV} emission lines in group four suggests that the combination of
projected rotation and outflow velocities is maximum at intermediate
inclinations.

\item
Since radio--quiet QSOs are associated with compact radio sources the
radio morphology cannot serve as an indicator of orientation.
Nevertheless, we notice that BALQSOs are preferentially found in group
two and accordingly we identify this group with highly inclined QSOs
so that the observer's line of sight passes through the dense, fast
wind. These are, in a sense, the radio--quiet analogs of the QSOs in
group five. Group two QSOs are UV weak and soft X--ray weak, which
suggests that they are are highly absorbed.  The high absorption fits
in very well with our proposed picture for the wind properties: it
appears only in radio--quiet QSOs where the wind is hypothesized to be
dense, but not in radio--loud QSOs, where it is not.

\item
Finally, group one can be taken to represent intermediate--to low
inclination cases among radio--quiet QSOs. This group contains the
radio--quiet analogs of group three and group four QSOs. The
relatively small number of QSOs in this group is likely the reason
that this group does not significantly break up via the
{\CIV~\fwhm}. As a result of the lack of an orientation indicator in
radio--quiet QSOs we cannot pick out a subset of such objects without
intrinsic NALs as could be done with their radio--loud
counterparts. Moreover, if the winds of radio--quiet QSOs are indeed
denser than those of radio--loud ones, there may be no viewing
direction that can avoid NAL gas completely.

\end{itemize}

The ability of intrinsic NALs to appear either blueward or redward of
the broad emission line (BEL) redshifts can be understood as a
consequence of the velocity field of the wind. The simulations of PSK
show that gas returning from a high altitude above the disk can be
entrained by the fast outflowing stream and flow inwards towards the
center. Thus it can produce absorption lines that are redshifted
relative to the peak of the broad emission lines. Alternatively, the
rotation of the wind can also result in redshifted NALs since the
projected velocity of the NAL gas can be either positive or negative,
depending on its azimuthal position.

A particularly interesting issue is raised by the fact that we find
only one strong associated NAL ($\rew($\CIV$) \gtrsim 1.5$~\AA) in our
low redshift sample. In comparison, previous surveys of
higher--redshift QSOs has revealed a considerably larger frequency of
such strong associated NALs in objects of comparable luminosity (see
\S6.1, above). This may be a consequence of an evolution of the
fueling rate of QSOs with redshift. If the QSO fueling rate is higher
at higher redshifts, then one might expect the mass loss rate, hence
the density of the wind, to be higher as well. Therefore, the simple
picture that we have examined offers the tantalizing prospect of
tracing the fueling history of QSOs through the strength of their
intrinsic NALs.

\acknowledgments

This research was funded by NASA through grant NAG 5--6399 and through
an archival award from Space Telescope Science Institute, which is
operated by AURA, Inc., under NASA contract NAS 5--26555. The authors
are grateful to Don Schneider, Buell Jannuzi, and Sofia Kirhakos for
their help and for providing the fully reduced Key Project data set,
including continuum fits. The authors would like to thank Fred Hamann
for an enlightening colloquium and several useful discussions. In
addition, RG would like to thank the technical support staff (in
particular, Mike Fowler and Katy Harrell) at Statsoft, Inc. who
responded to the incessant questioning regarding the Statistica
multivariate analysis package and the application of Ward's
amalgamation method. WNB acknowledges the support of NSF CAREER award
AST--9983783.


\clearpage
\begin{figure}
\figurenum{1}
\plotone{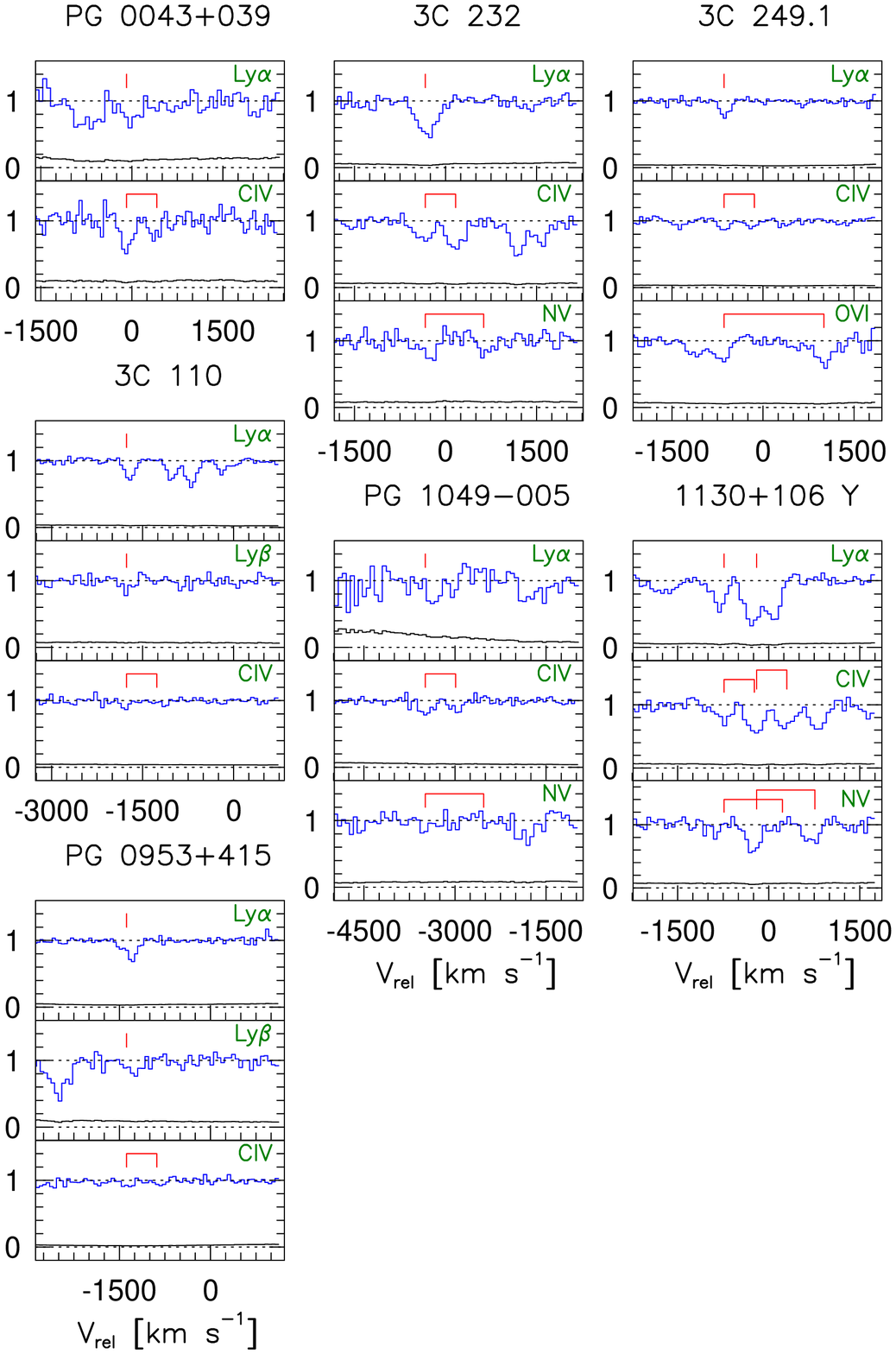}
\vglue -0.7in
\figcaption[fig1a.col.eps]{
\scriptsize
{\bf Velocity--aligned plots:} For each NALQSO candidate, we show the
spectrum of all detected ions (at a {$3\sigma$} threshold) in
increasing atomic number and aligned in velocity with respect to the
QSO emission redshift. A positive velocity indicates wavelengths
redward of the emission redshift. Note that the broad emission lines
are usually blueshifted relative to the systemic redshift.
Deviations in the wavelength calibrations across the FOS gratings
result in small velocity shifts between ions that appear on different
gratings. We note that the FOS wavelength scale zero point is
repeatable to 80~\kms at \Lya.}
\label{fig:systems}
\end{figure}

\clearpage

\begin{figure}
\figurenum{1}
\plotone{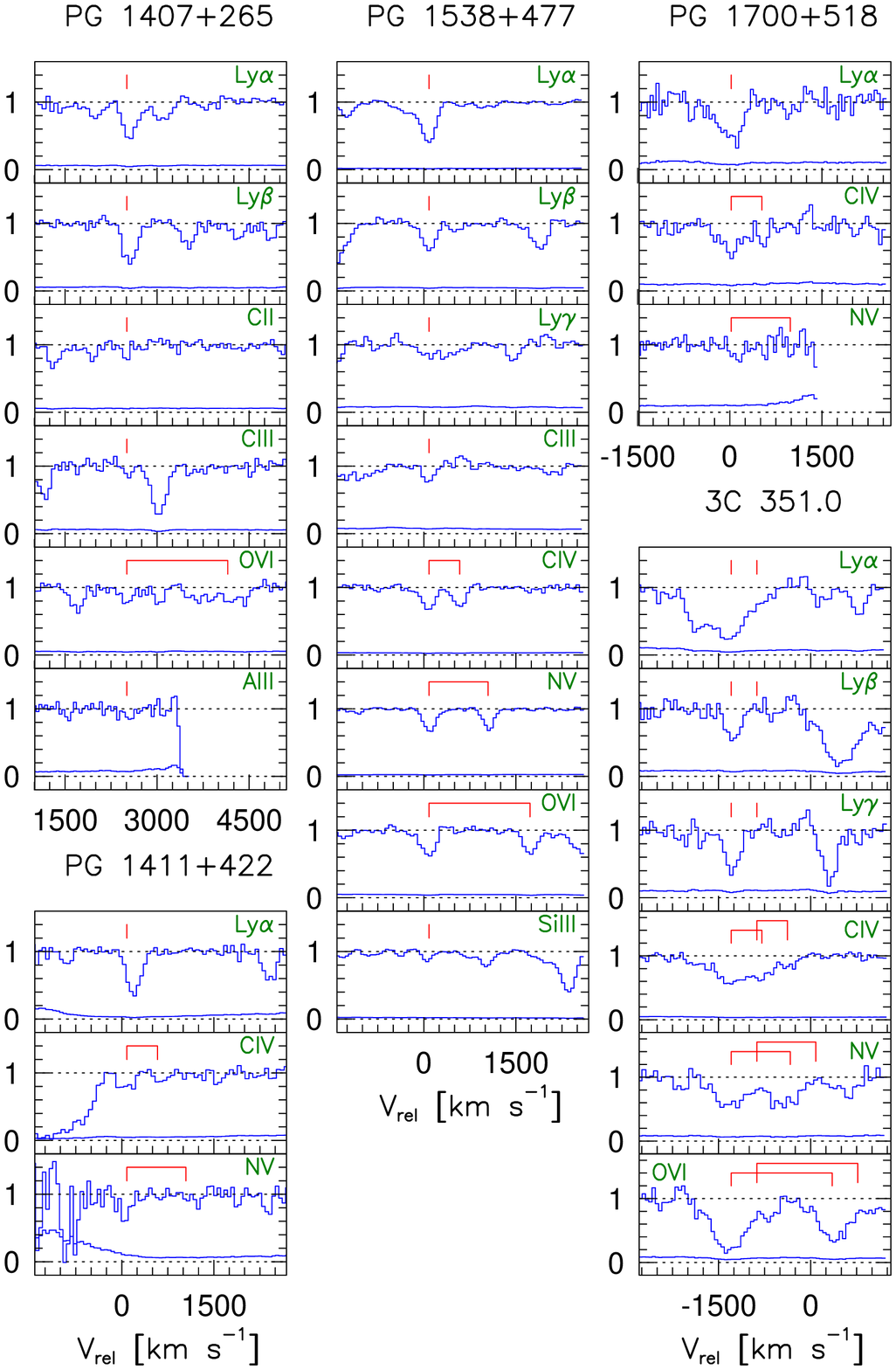}
\vglue -0.7in
\figcaption[fig1b.col.eps]{
\scriptsize
continued}
\end{figure}

\clearpage

\begin{figure}
\figurenum{1}
\plotone{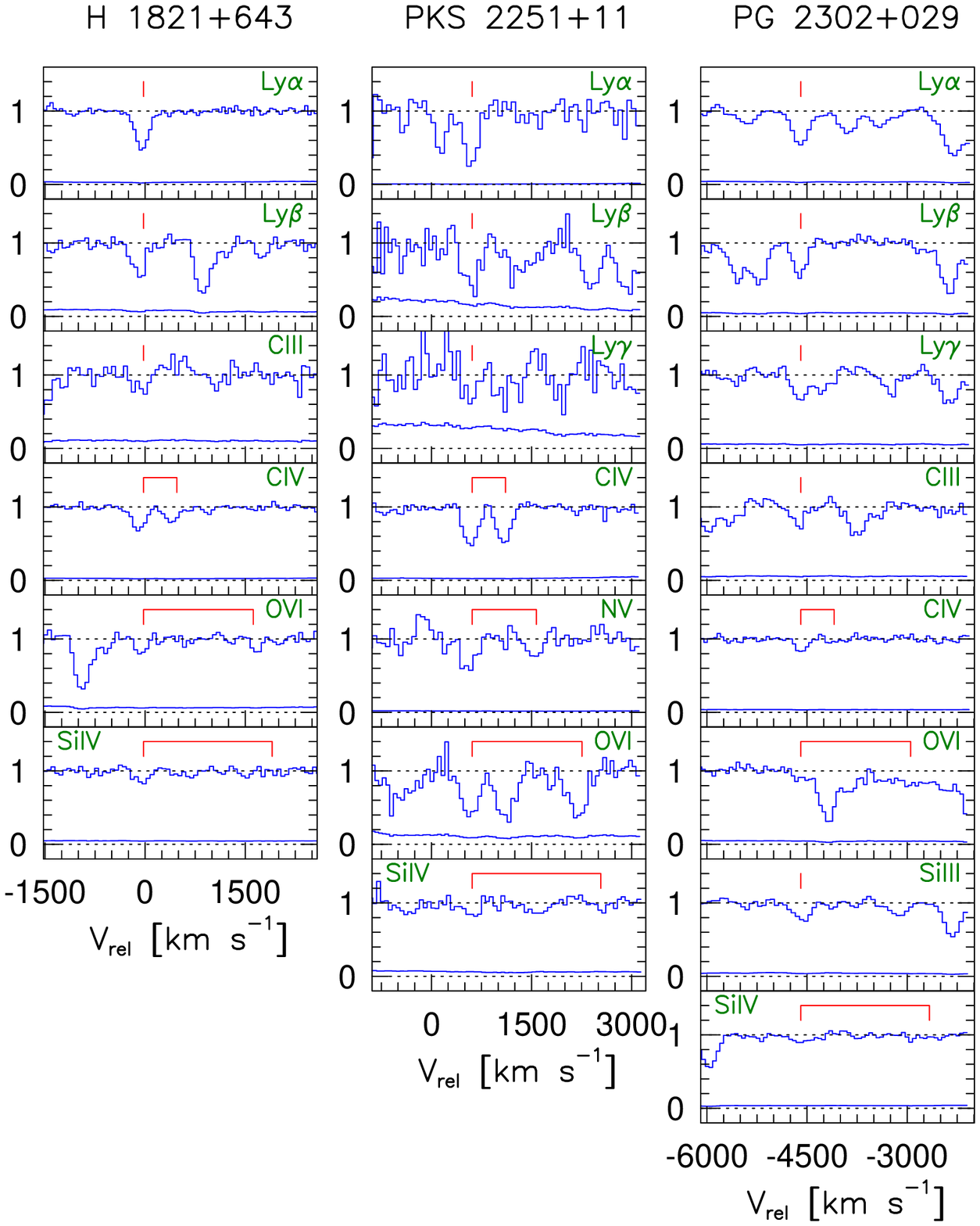}
\vglue -0.7in
\figcaption[fig1c.col.eps]{
\scriptsize
continued}
\end{figure}

\clearpage

\begin{figure}[th]
\figurenum{2}
\plotone{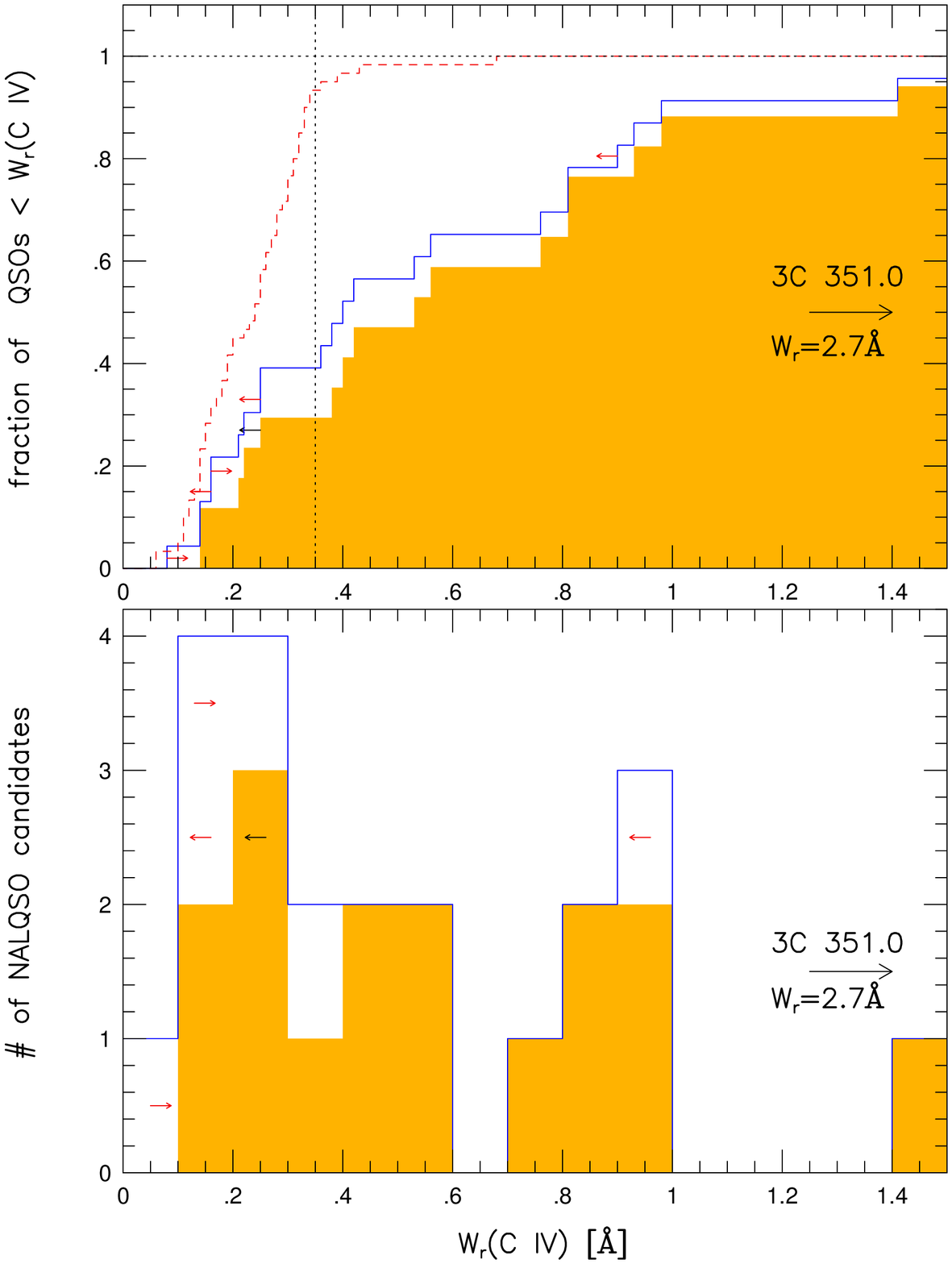}
\vglue -0.7in
\figcaption[fig2.col.eps]{
\scriptsize
{\bf Equivalent Width Distributions:} In the bottom panel, we show the
distribution of {C\,{\sc iv}} rest--frame equivalent widths ($\rew$) of the
$\lambda$1548 transition for the $\alpha$-- (shaded) and the $\beta$--
(unshaded) samples.  In the top panel, we show the cumulative
distribution of {$\rew($C\,{\sc iv}$)$} for the {$\alpha$} (shaded) and
{$\beta$} (unshaded) samples and the {$3\sigma~\rew$} limit
(dashed). Arrows indicate limits on {C\,{\sc iv}}.  The vertical dotted line
demarks the 95\% sensitivity cutoff at {$\rew=0.35$~\AA}.}
\label{fig:ewhisto}
\end{figure}

\clearpage

\begin{figure}[th]
\figurenum{3}
\plotone{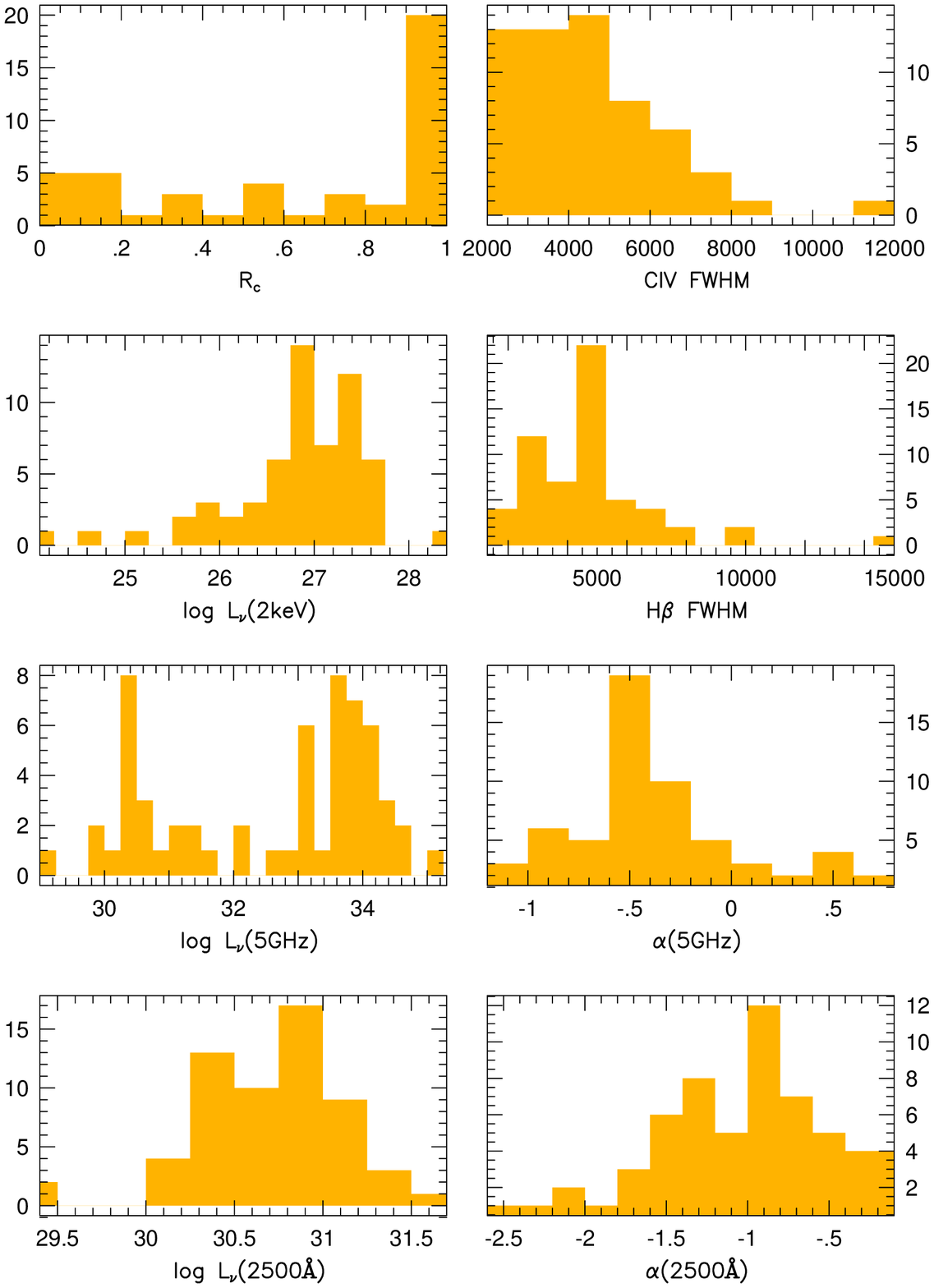}
\vglue -0.7in
\figcaption[fig3.col.eps]{
\scriptsize
{\bf Distributions of QSOs:} For the eight measured properties ({$\log
L_\nu(\mathrm{2500}$~\AA$)$}, {$\log L_\nu(\mathrm{5~GHz})$}, {$\log
L_\nu(\mathrm{2~keV})$} {$\alpha(\mathrm{2500}$~\AA$)$},
{$\alpha(\mathrm{5~GHz})$}, {C\,{\sc iv~fwhm}}, {H$\beta$~{\sc
fwhm}}, {$R_{\mathrm{c}}$} ), we show the distributions of the QSOs in
the {$\alpha$}--sample.}
\label{fig:distrib}
\end{figure}

\begin{figure}[th]
\figurenum{4}
\plotone{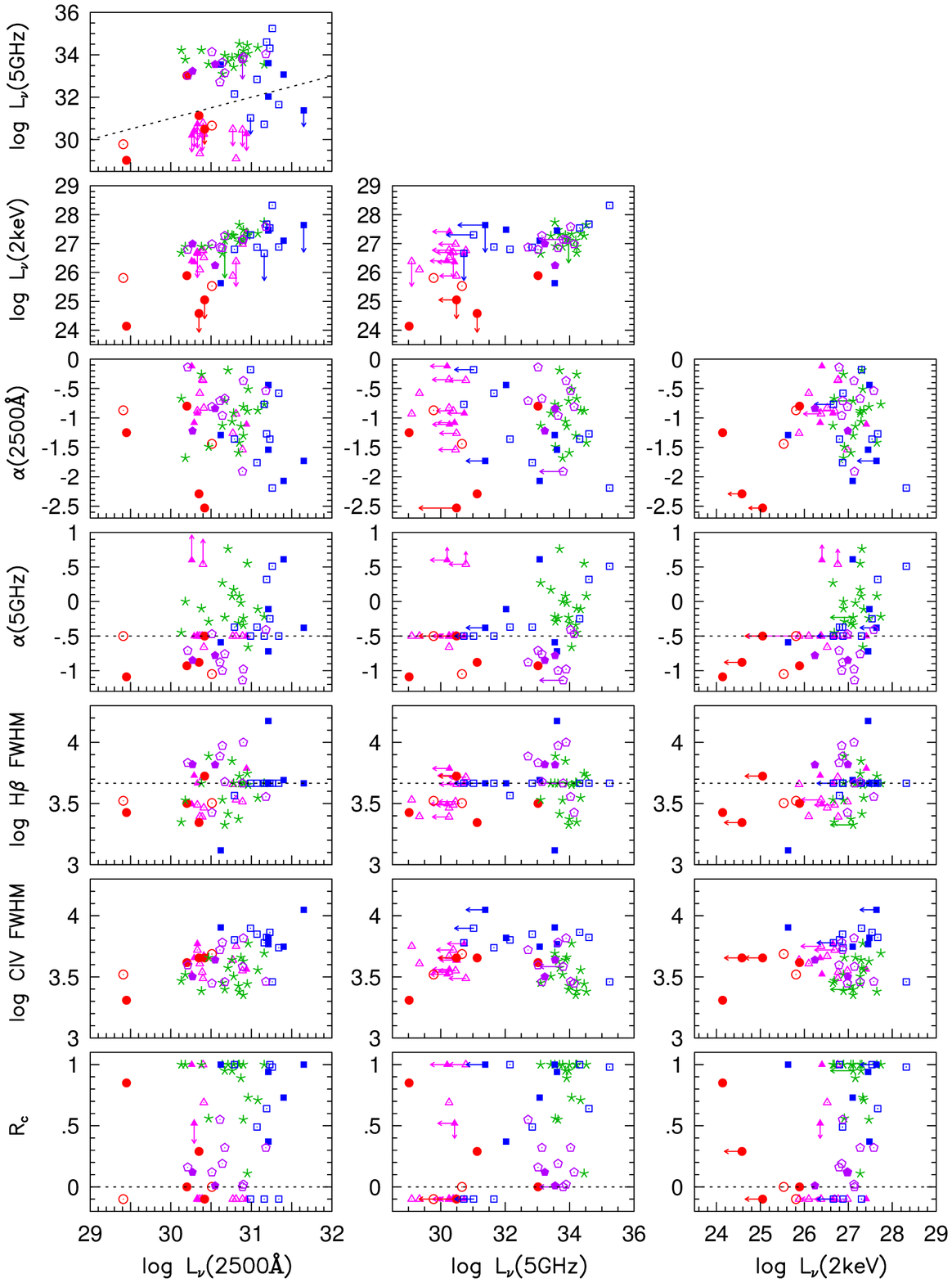}
\vglue -0.7in
\figcaption[fig4a]{
\scriptsize
{\bf Scatter Plots of QSO properties:} For all measured properties of
the QSOs in the sample, we plot all combinations of bivariate scatter
plots.  Units are the same as in Table~3. NALQSO candidates are
represented by filled symbols while ``normal'' QSOs are shown as open
symbols. The symbol shapes correspond to the five groups of QSOs as
defined by the multivariate analysis (group one -- triangle; group two
-- circles; group three -- stars; group four -- boxes; group five --
pentagons). There are no NALQSO candidates in group three. The dashed
line in the {$\log L_\nu(\mathrm{5~GHz})~\mathrm{vs.}~\log
L_\nu($2500~\AA$)$} plot marks the division between radio--loud
(top half) and radio--quiet (bottom half) QSOs. In plots involving
{$\alpha(\mathrm{5~GHz})$} or {$\log \mathrm{H}\beta$~\fwhm}, the
dashed line indicates the assumed value when no measurement was
available [{$\alpha(\mathrm{5~GHz})=-0.5$},
{H$\beta$~{\fwhm}$=4630~\mathrm{km~s}^{-1}$}].}
\label{fig:scatter}
\end{figure}

\clearpage

\begin{figure}[th]
\figurenum{4}
\plotone{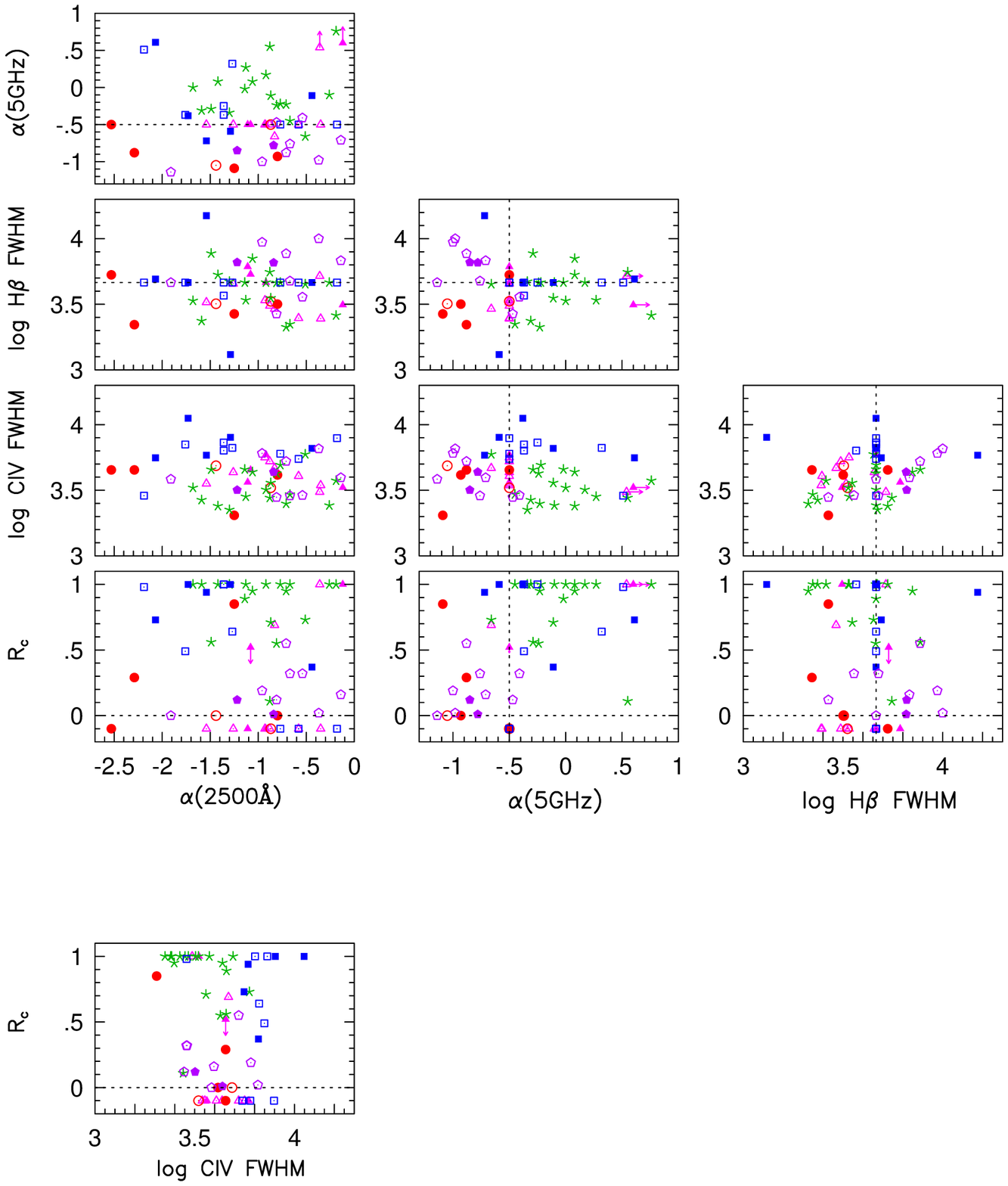}
\vglue -0.7in
\figcaption[fig4b.col.eps]{
\scriptsize
continued}
\end{figure}

\clearpage

\begin{figure}[th]
\figurenum{5}
\plotone{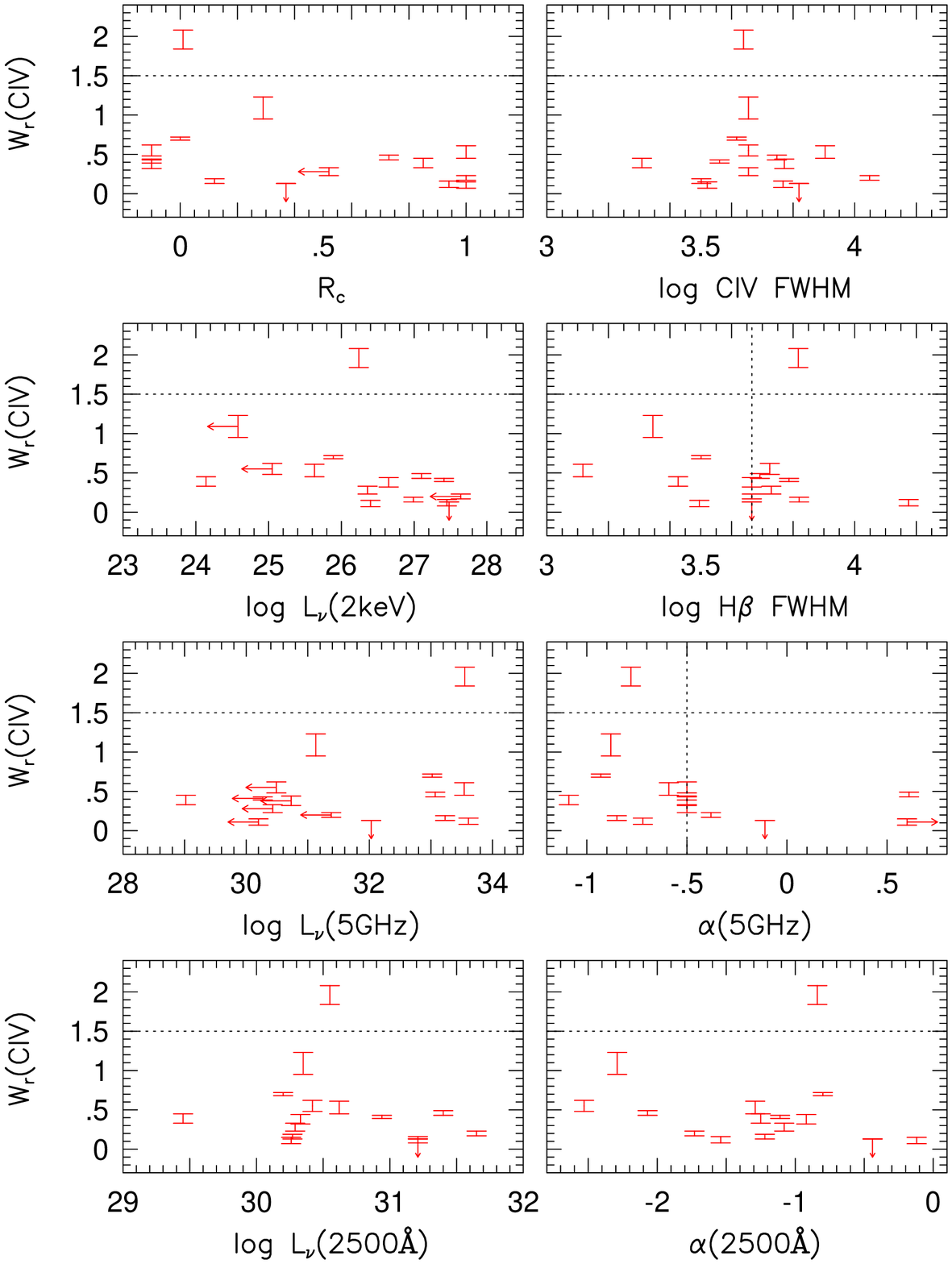}
\vglue -0.7in
\figcaption[fig5.col.eps]{
\scriptsize
{\bf Equivalent Width Mosaic:} We plot the {C\,{\sc iv}} NAL equivalent width
against the eight measured QSO properties for the NALQSO candidates. Units
are as reported in Table~2 and Table~3. Vertical lines represent assumed
values (see Fig.~3). The horizontal dashed line marks the Foltz \etal~(1986)
division between strong and weak NALs. Our low redshift sample only has one
strong NAL.}
\label{fig:ewmosaic}
\end{figure}

\clearpage

\begin{figure}[th]
\figurenum{6}
\plotone{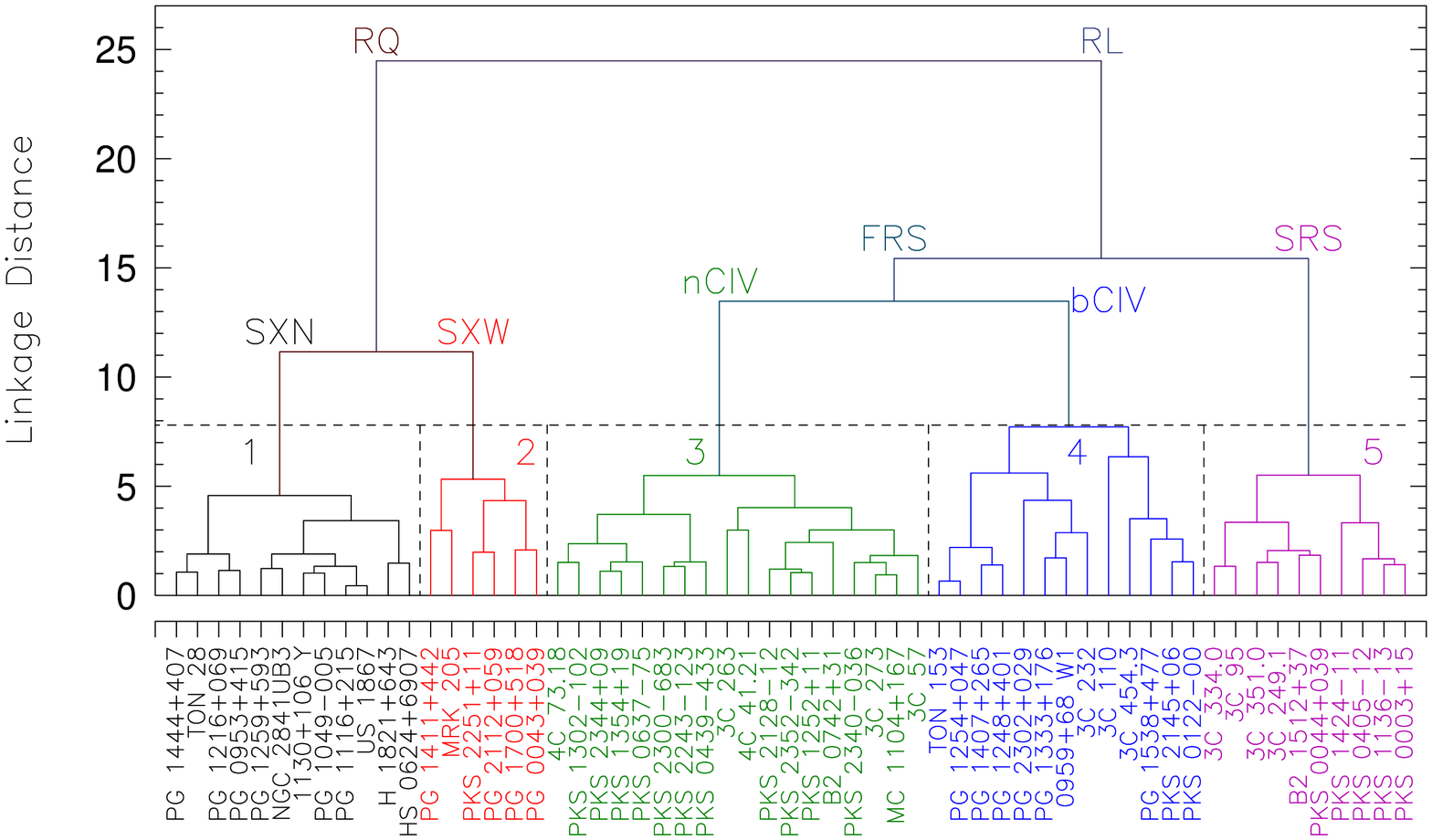}
\vglue -0.8in
\figcaption[fig6.col.eps]{
\scriptsize
{\bf Amalgamation Schedule of QSOs:} The QSOs group together via the
schedule shown. (See the text for a description of the grouping
procedure.)  Ward's method was used to determine the amalgamation
schedule for the 59 QSOs, and distances were computed using a
Euclidean metric. The positions of QSOs in parameter space relied on
standardized coordinates as described in the text. The codes are:
RQ$=$radio--quiet; RL$=$radio--loud; FSR$=$flat radio spectrum;
SRS$=$steep radio spectrum; nCIV$=$normal/narrow {C\,{\sc iv}}
emission line {\fwhm}; bCIV$=$broad {C\,{\sc iv}} emission line
{\fwhm}; SXN$=$soft X--ray ``normal''; SXW$=$soft X--ray weak.  The
most distinct groupings occur at distances (sizes) {$<$7.8}. The
second group contains all the BALQSOs (except {PG~$1254+047$}) in the
sample while the third group is unique in that it contains no NALQSO
candidates.}
\label{fig:cluster}
\end{figure}

\clearpage

\begin{figure}[th]
\figurenum{7}
\plotone{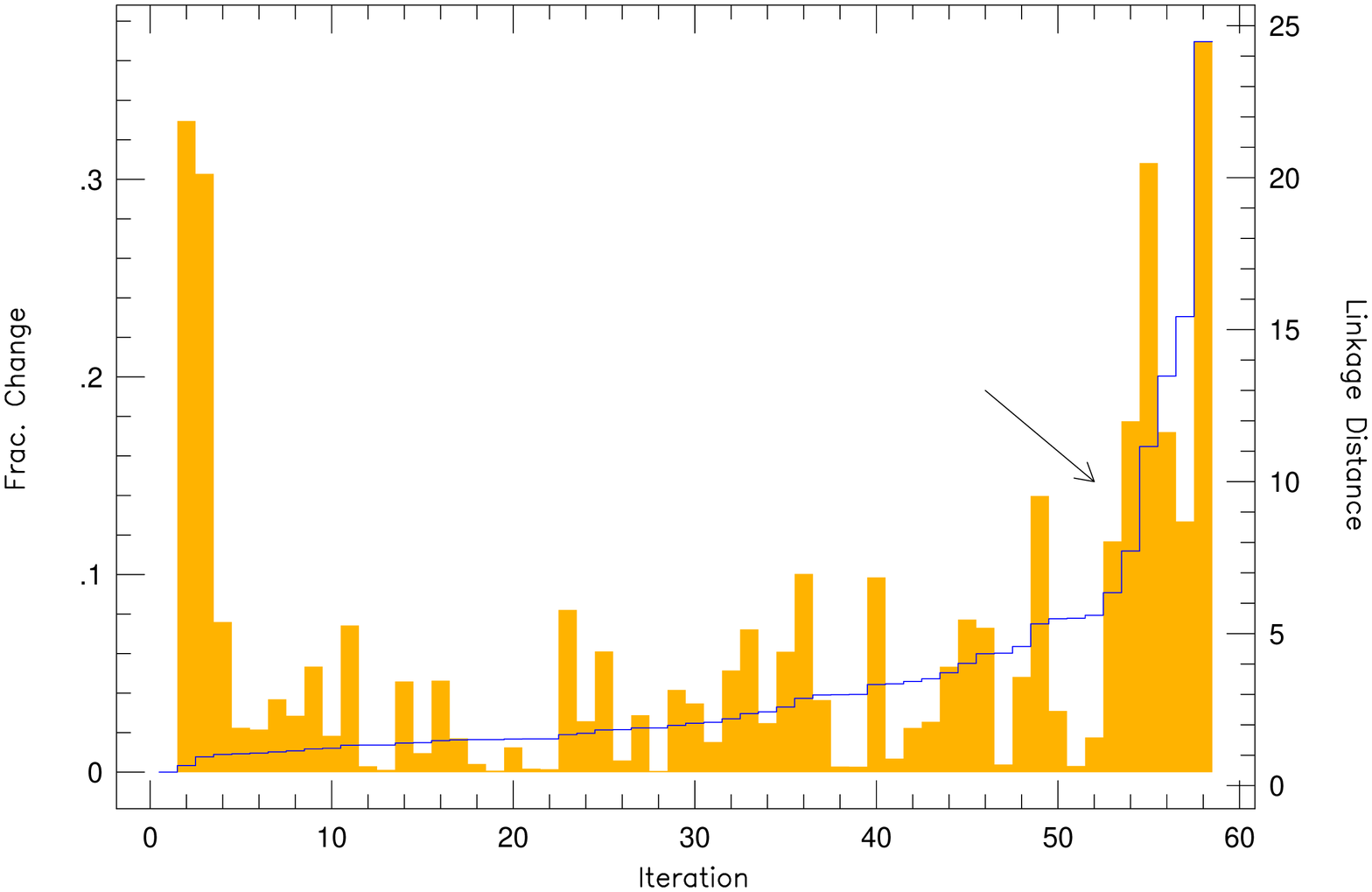}
\vglue -0.6in
\figcaption[fig7.col.eps]{
\scriptsize
{\bf Scree Plot:} For each iteration of the amalgamation schedule,
we plot the linkage distance (line) and the fractional change in the distance
(shaded histogram). Ignoring the first and last few iterations, the largest
fractional change occurs when the distance jumps from {$\sim7.8$} to
{$\sim11.2$} (arrow). Groups with sizes less than {$\sim7.8$} are not
significantly distinct from their nearest neighbors.}
\label{fig:scree}
\end{figure}

\clearpage

\begin{figure}[th]
\figurenum{8}
\plotone{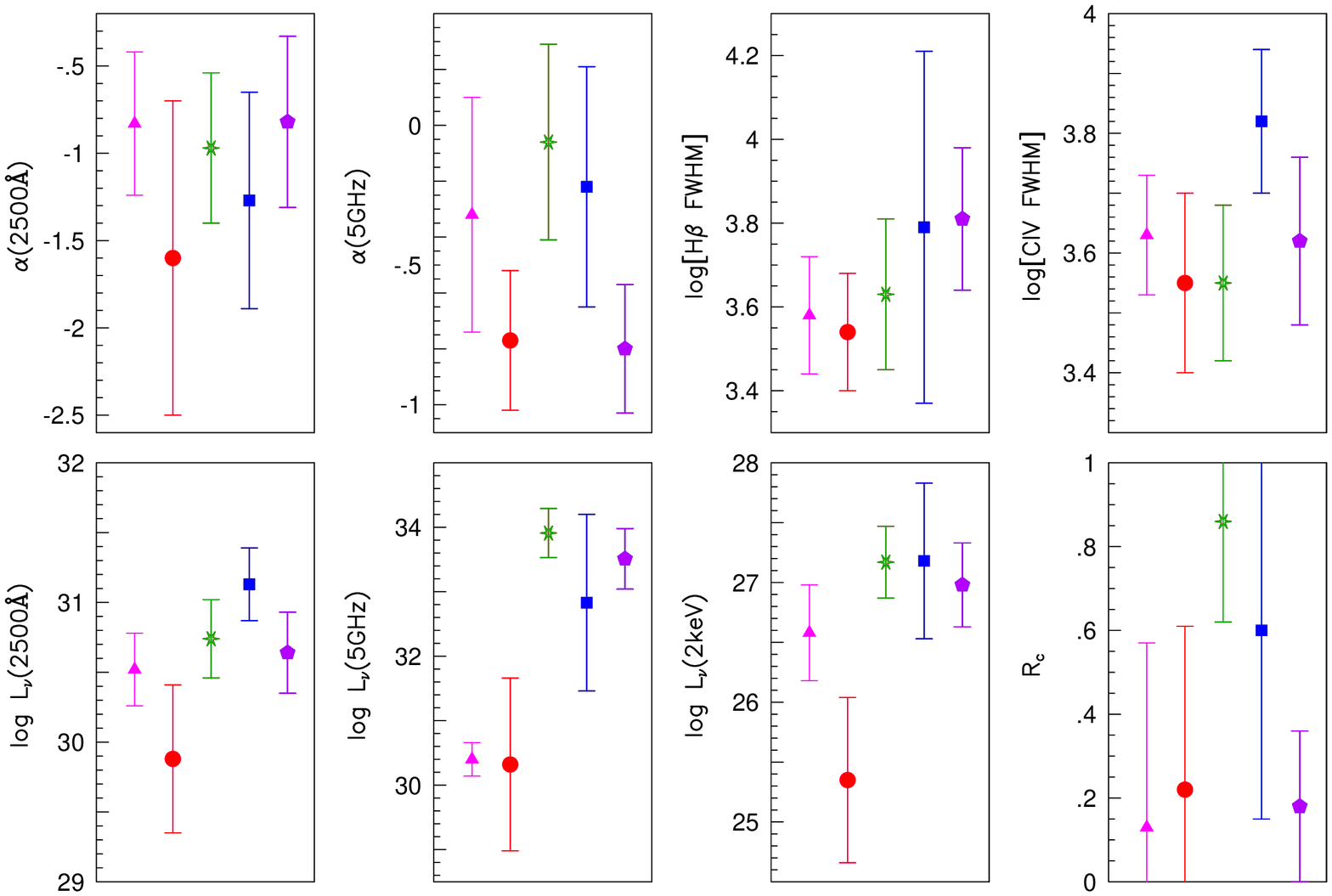}
\vglue -0.7in
\figcaption[fig8.col.eps]{
\scriptsize
{\bf Mean Properties of Clusters:} For each property used in the
clustering analysis, we show the mean and {$1\sigma$} standard
deviations for each of the five groups (group one -- triangle; group
two -- circle; group three -- star; group four -- box; group five --
pentagon). The units are as in Table~3. Group three has 18 of the 59
QSOs in the sample, none of which are candidate NALQSOs. This group is
distinguished by a high 5~GHz luminosity density, high radio core
fraction, flat radio spectral index and a mediocre {C\,{\sc iv~fwhm}}
($<$6000~\kms).}
\label{fig:kmean}
\end{figure}

\clearpage

\begin{figure}[th]
\figurenum{9}
\plotone{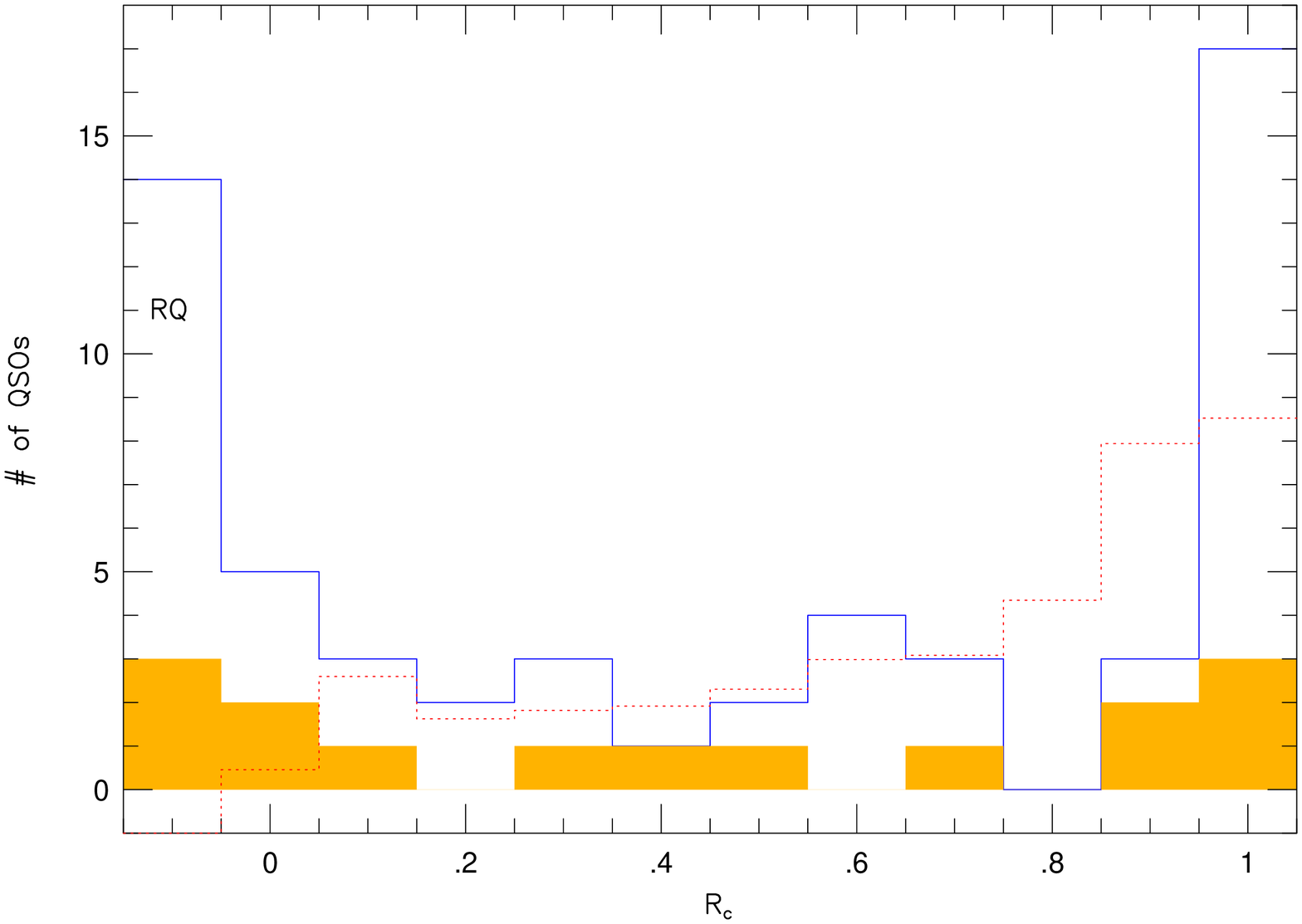}
\vglue -0.6in
\figcaption[fig9.col.eps]{
\scriptsize
{\bf Histogram of Radio Core Fractions:} The radio core fraction
distribution is shown for all QSOs (unshaded histogram) in the
sample. The NALQSO candidates are shown by the shaded histogram. In
cases where the core fraction could not be found or measured, one was
assigned based on the radio spectral index ({$R_{\mathrm{c}}\sim1$}
when {$\alpha>>-0.5$} and {$R_{\mathrm{c}}\sim0$} when
{$\alpha<<-0.5$}).  Radio--quiet QSOs for which a core fraction could
not be found, measured, or assigned based on a strongly constrained
spectral index have been placed in the unphysical bin at
{$R_{\mathrm{c}}=-0.1$}. Ignoring this bin, the distribution is
consistent with a sample of randomly oriented accretion disks, if
{$R_{\mathrm{c}} \sim \cos i$}, where {$i$} is the angle between the
observer and the normal to the disk.}
\label{fig:rchisto}
\end{figure}

\clearpage

\begin{figure}[th]
\figurenum{10}
\plotone{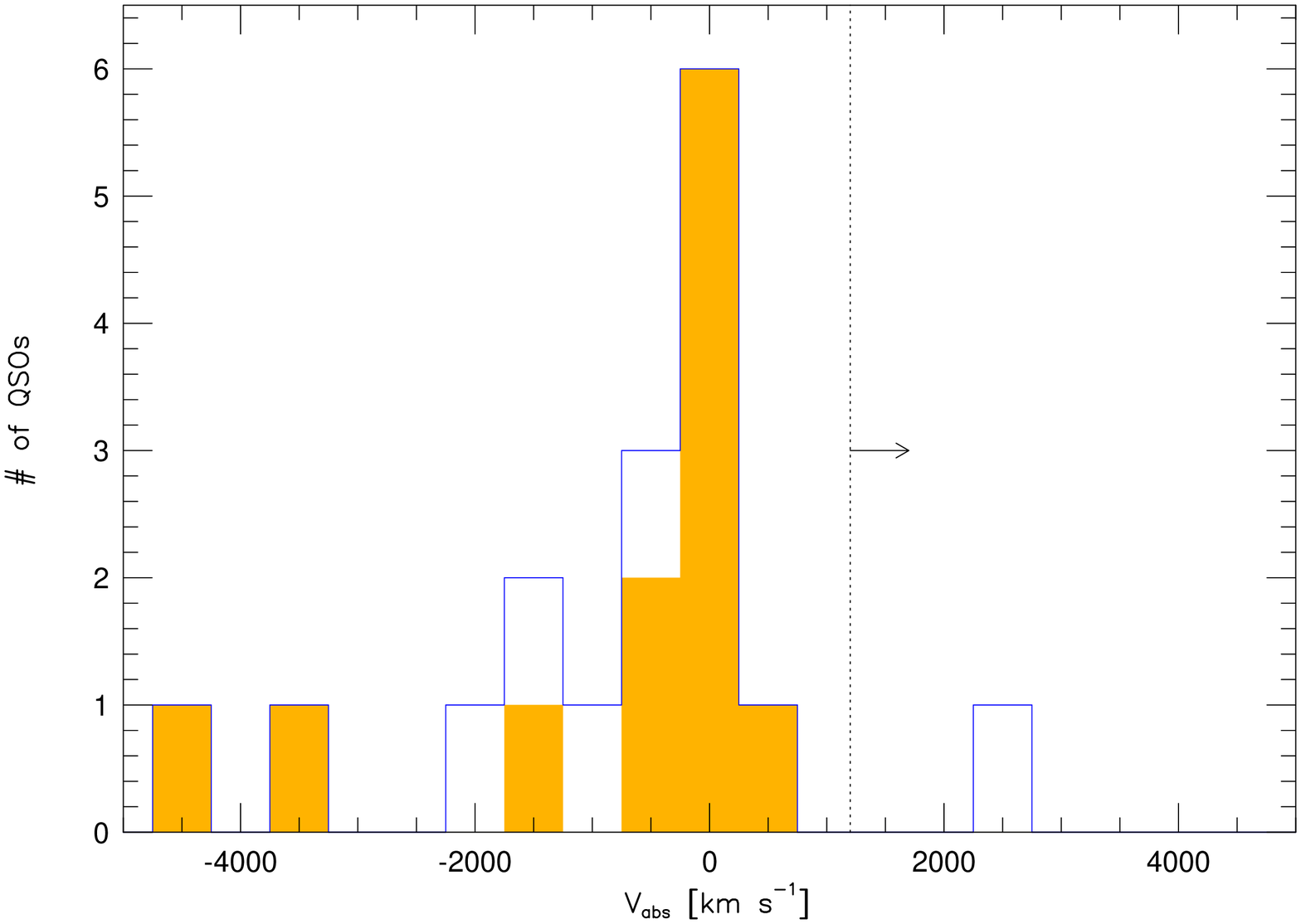}
\vglue -0.6in
\figcaption[fig10.col.eps]{
\scriptsize
{\bf Histogram of Absorber Velocities:} We show a velocity histogram
of the narrow absorption line systems. Velocities were computed
relative to the QSO emission redshift with positive velocities redward
of the emission redshift. The unshaded histogram is for all 17
systems, while the shaded histogram is for the
{$\rew>0.35$~\AA} systems, for which we are 95\%
complete. While the systemic redshifts of the QSOs in the sample are
not known (finding them is beyond the scope of the paper), the dotted
line crudely demarks the ``systemic'' velocity which is typically more
than {1200~\kms} redward of the UV emission redshift. The velocity
distribution of the narrow absorption lines is roughly centered on the
QSO emission redshift.}
\label{fig:vejhisto}
\end{figure}

\clearpage

\begin{figure}[th]
\figurenum{11}
\plotone{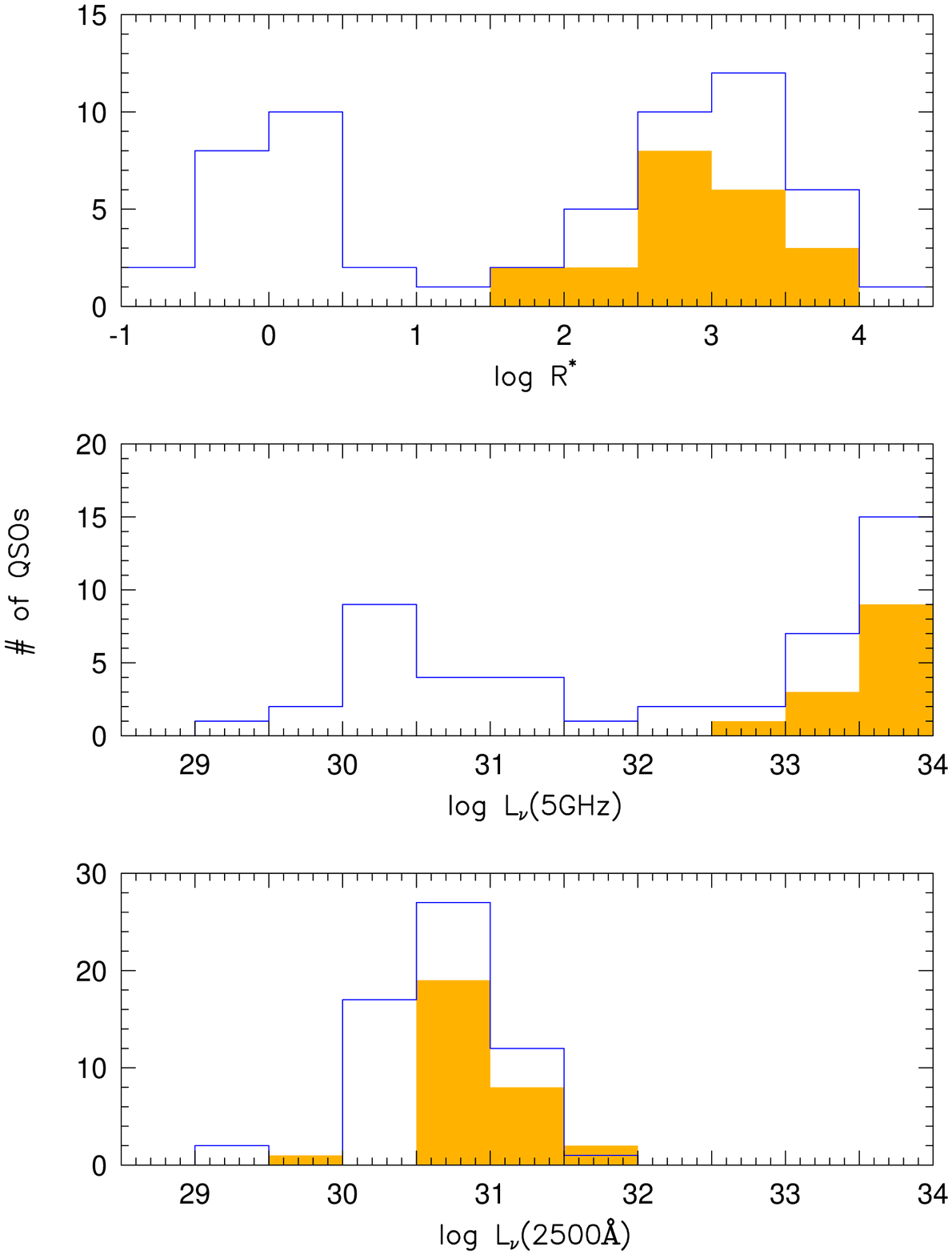}
\vglue -0.7in
\figcaption[fig11.col.eps]{
\scriptsize
{\bf Comparison of the Foltz \etal~(1986) sample with the
$\alpha$--sample:} We compare the optical and radio luminosity
distributions of the 21 QSOs in the Foltz \etal~(1986) sample (shaded)
with the 59 QSOs in the {$\alpha$}--sample (unshaded).  Our QSOs
clearly sample larger portions of the luminosity functions. However, a
subsample of our QSOs which overlaps the Foltz \etal~(1986) sample
shows no strong narrow absorption lines. Thus, there may be an
evolution of strong associated systems from intermediate to low
redshift.}
\label{fig:foltzcomp}
\end{figure}

\clearpage

\begin{figure}[th]
\figurenum{12}
\plotone{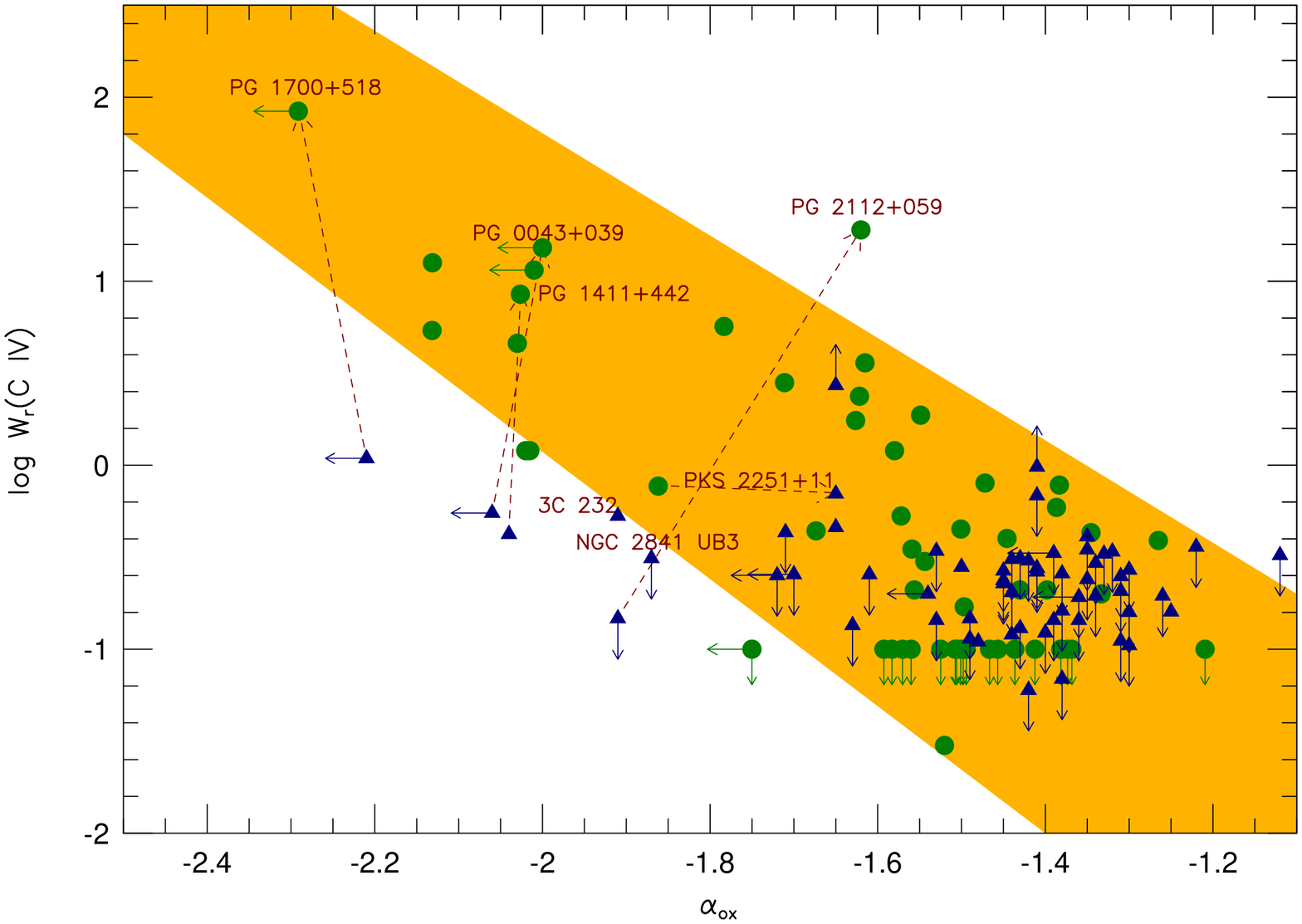}
\vglue -0.6in
\figcaption[fig12.col.eps]{
\scriptsize
{\bf {C\,{\sc iv}} Equivalent Width vs. Optical/X--ray Spectral Index:}
Reproduced from BLW, we plot the {C\,{\sc iv}$\lambda$1548} absorption
equivalent width (or limit) for each QSO against the optical/X--ray
spectral index. The triangles are the QSOs in the {$\alpha$}--sample
and the circles are the {$z_{\mathrm{em}}<0.5$} Palomar--Green QSOs
from the BLW sample. The shaded region is meant to suggest the general
anti--correlation described by BLW. With one exception
({PKS~$2251+11$}), the dotted lines are corrections to the
{$\alpha$}--sample BALQSOs based on the equivalent width of the BAL
gas and improvements in {$\aox$} using {\it ASCA} data (Gallagher
\etal~2000). The dashed line for {PKS~$2251+11$} connects the BLW point
with our point.}
\label{fig:niel}
\end{figure}

\clearpage

\begin{figure}[th]
\figurenum{13}
\plotone{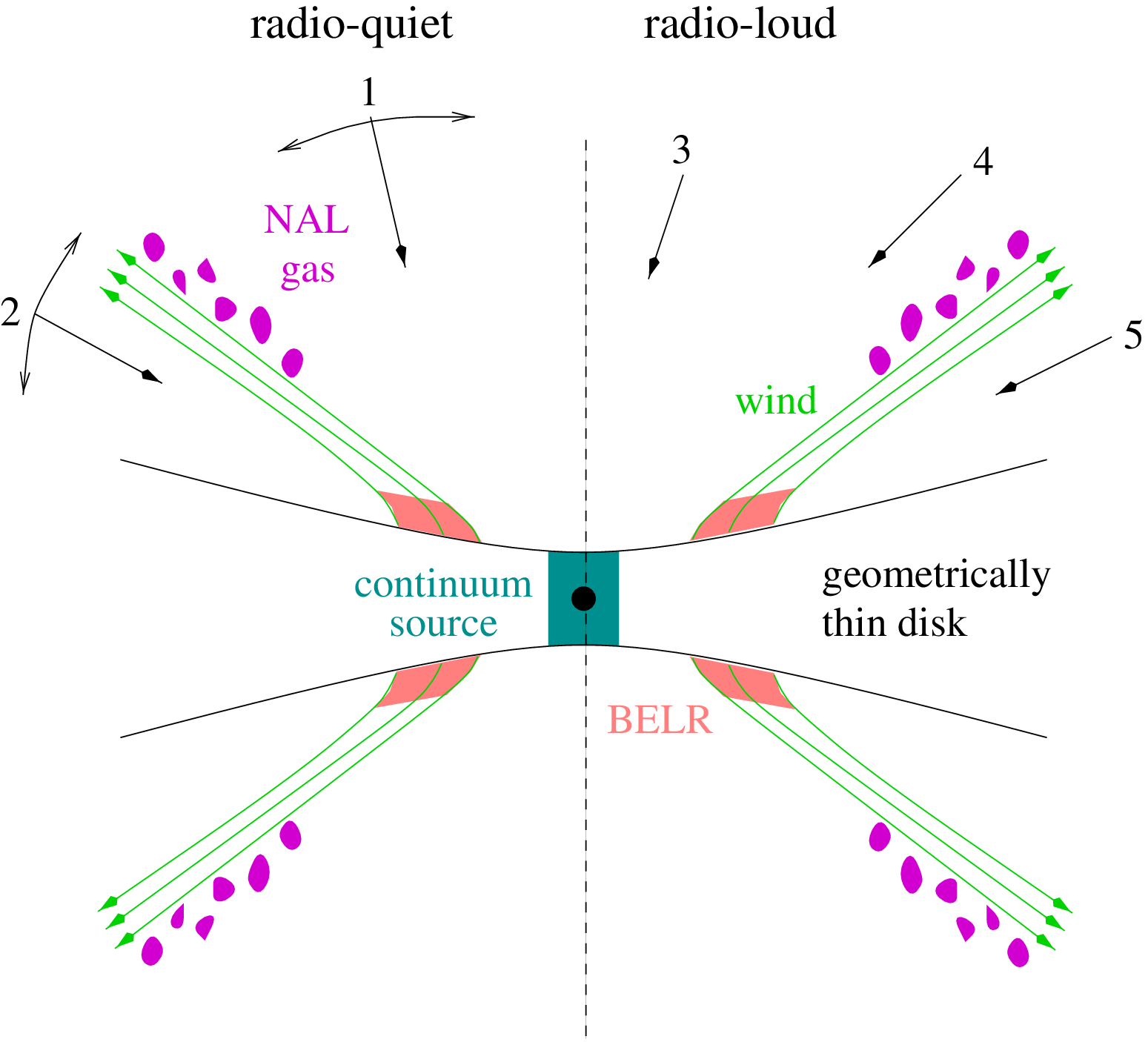}
\figcaption[fig13.col.eps]{
\scriptsize
{\bf Cartoon of Disk--Wind Model for QSOs:} This is a schematic
illustration of the disk--wind scenario of MCGV which explained the
BALQSO phenomenon as an orientation effect. The central region of the
accretion disk (dark shade) is the source of continuum UV and X--ray
photons, while the inner portion of the wind (light shade) is the
source of UV high--ionization line emission (e.g., {C\,{\sc iv}},
{N\,{\sc v}}) and {\Lya}. In this picture, NAL gas is viewed as a
clumpy medium that ``hugs'' the wind relatively far from the inner
accretion disk. The five groups of QSOs can be explained through a
combination of inclination angle, $i$, and radio--loudness, as
described in \S7.2 of the text.}
\label{fig:diskwind}
\end{figure}

\clearpage

\begin{deluxetable}{lccclccc}
\tabletypesize{\tiny}
\tablewidth{0pc}
\tablenum{1}
\tablecaption{Subsample of Key Project QSOs}
\tablehead
{
\colhead{Object} &
\colhead{$z_{\mathrm{em}}$\tablenotemark{a}} &
\colhead{$V$\tablenotemark{a}} &
\colhead{Sample\tablenotemark{b}} &
\colhead{Object} &
\colhead{$z_{\mathrm{em}}$\tablenotemark{a}} &
\colhead{$V$\tablenotemark{a}} &
\colhead{Sample\tablenotemark{b}} \\
\colhead{} &
\colhead{} &
\colhead{(mag)} & 
\colhead{} &
\colhead{} &
\colhead{} &
\colhead{(mag)} &
\colhead{}
}
\startdata
{PKS~$0003+15$}  & 0.450 & 16.5 & $\alpha$ & {PKS~$1252+11$}  & 0.870 & 16.6 & $\alpha$ \\
{PG~$0043+039$}  & 0.384 & 15.8 & $\alpha$ & {PG~$1254+047$}  & 1.024 & 16.4 & $\alpha$ \\
{PKS~$0044+030$} & 0.624 & 16.1 & $\alpha$ & {PG~$1259+593$}  & 0.472 & 15.7 & $\alpha$ \\
{PKS~$0122-00$}  & 1.070 & 16.7 & $\alpha$ & {PG~$1302-102$}  & 0.286 & 15.2 & $\alpha$ \\
{3C~$57$}        & 0.670 & 16.4 & $\alpha$ & {TON~$153$}      & 1.022 & 16.0 & $\alpha$ \\
{3C~$95$}        & 0.614 & 16.2 & $\alpha$ & {PG~$1333+176$}  & 0.554 & 16.1 & $\alpha$ \\
{PKS~$0405-12$}  & 0.574 & 14.9 & $\alpha$ & {PKS~$1354+19$}  & 0.720 & 16.0 & $\alpha$ \\
{3C~$110$}       & 0.773 & 16.0 & $\alpha$ & {PG~$1407+265$}  & 0.940 & 16.0 & $\alpha$ \\
{PKS~$0439-433$} & 0.593 & 16.4 & $\alpha$ & {PG~$1411+442$}  & 0.089 & 15.0 & $\alpha$ \\
{HS~$0624+6907$} & 0.370 & 14.2 & $\alpha$ & {PG~$1415+451$}  & 0.114 & 15.7 & $\alpha$ \\
{PKS~$0637-75$}  & 0.654 & 15.8 & $\alpha$ & {PKS~$1424-11$}  & 0.805 & 16.5 & $\alpha$ \\
{B2~$0742+31$}   & 0.462 & 16.4 & $\alpha$ & {PG~$1444+407$}  & 0.267 & 15.7 & $\alpha$ \\
{US~$1867$}      & 0.513 & 16.4 & $\alpha$ & {B2~$1512+37$}   & 0.371 & 15.5 & $\alpha$ \\
{NGC~$2841$~UB3} & 0.553 & 16.5 & $\alpha$ & {PG~$1538+477$}  & 0.770 & 16.0 & $\alpha$ \\
{PG~$0953+415$}  & 0.239 & 15.3 & $\alpha$ & {3C~$334.0$}     & 0.555 & 16.4 & $\alpha$ \\
{3C~$232$}       & 0.533 & 15.8 & $\alpha$ & {PG~$1700+518$}  & 0.290 & 15.4 & $\alpha$ \\
{$0959+68$~W1}   & 0.773 & 15.9 & $\alpha$ & {3C~$351.0$}     & 0.371 & 15.3 & $\alpha$ \\
{TON~$28$}       & 0.329 & 16.0 & $\alpha$ & {H~$1821+643$}   & 0.297 & 14.2 & $\alpha$ \\
{4C~$41.21$}     & 0.613 & 16.9 & $\alpha$ & {4C~$73.18$}     & 0.302 & 15.5 & $\alpha$ \\
{PG~$1049-005$}  & 0.357 & 16.0 & $\alpha$ & {PG~$2112+059$}  & 0.457 & 15.7 & $\alpha$ \\
{3C~$249.1$}     & 0.311 & 15.7 & $\alpha$ & {PKS~$2128-12$}  & 0.501 & 16.1 & $\alpha$ \\
{MC~$1104+167$}  & 0.634 & 15.7 & $\alpha$ & {PKS~$2145+06$}  & 0.990 & 16.5 & $\alpha$ \\
{PG~$1116+215$}  & 0.177 & 15.0 & $\alpha$ & {PKS~$2243-123$} & 0.630 & 16.5 & $\alpha$ \\
{$1130+106$~Y}   &  0.51 & 17.0 & $\alpha$ & {3C~$454.3$}     & 0.859 & 16.1 & $\alpha$ \\
{PKS~$1136-13$}  & 0.554 & 16.1 & $\alpha$ & {PKS~$2251+11$}  & 0.323 & 15.8 & $\alpha$ \\
{3C~$263$}       & 0.652 & 16.3 & $\alpha$ & {PKS~$2300-683$} & 0.512 & 16.4 & $\alpha$ \\
{PG~$1202+281$}  & 0.165 & 15.6 & $\beta$  & {PG~$2302+029$}  & 1.052 & 15.8 & $\alpha$ \\
{PG~$1216+069$}  & 0.334 & 15.8 & $\alpha$ & {PKS~$2340-036$} & 0.896 & 16.0 & $\alpha$ \\
{Mrk~$205$}      & 0.071 & 14.5 & $\alpha$ & {PKS~$2344+09$}  & 0.677 & 16.0 & $\alpha$ \\
{3C~$273$}       & 0.158 & 12.8 & $\alpha$ & {PKS~$2352-342$} & 0.702 & 16.4 & $\alpha$ \\
{PG~$1248+401$}  & 1.030 & 16.1 & $\alpha$ \\
\enddata
\tablenotetext{a}{Emission redshifts and {$V$} magnitudes were taken from
Jannuzi \etal~(1998).}
\tablenotetext{b}{All QSOs in the {$\alpha$}--sample are also part of the
{$\beta$}--sample.}
\label{tab:sample}
\end{deluxetable}

\clearpage

\begin{deluxetable}{lrcccc}
\tabletypesize{\scriptsize}
\tablewidth{0pc}
\tablenum{2}
\tablecaption{Absorption Properties of NALQSO Candidates}
\tablehead
{
\colhead{Object} &
\colhead{$\Delta v_{\mathrm abs}$\tablenotemark{a}} &  
\colhead{$W_{\mathrm{r}}($C\,{\sc iv}$)$\tablenotemark{b}} &
\colhead{$W_{\mathrm{r}}($N\,{\sc v}$)$\tablenotemark{b}} &
\colhead{$W_{\mathrm{r}}($O\,{\sc vi}$)$\tablenotemark{b}} &
\colhead{$W_{\mathrm{r}}($Ly\,$\alpha)$} \\
\colhead{} &
\colhead{\kms} &
\colhead{\AA} &
\colhead{\AA} &
\colhead{\AA} &
\colhead{\AA}
}
\startdata
{PG~$0043+039$} &   $-90$ & $0.55\pm0.07$ & $<$0.27       & \nodata       & $0.47\pm0.10$ \\
{3C~$110$}      & $-1760$ & $0.12\pm0.04$ & $<$0.05       & $<0.10$       & $0.25\pm0.02$ \\
{PG~$0953+415$} & $-1380$ & $0.11\pm0.04$ & $<$0.09\tablenotemark{e}
                                                          & $<0.12$       & $0.29\pm0.02$ \\
{3C~$232$}      &  $-330$ & $0.53\pm0.08$ & $0.28\pm0.06$ & \nodata       & $0.83\pm0.04$ \\
{PG~$1049-005$} & $-3490$ & $0.28\pm0.05$ & $0.14\pm0.06$ & \nodata       & $0.35\pm0.13$ \\
{3C~$249.1$}    &  $-640$ & $0.16\pm0.03$ & \nodata       & $0.30\pm0.05$ & $0.21\pm0.02$ \\
{$1130+106$~Y}\tablenotemark{c}
                &  $-740$ & $0.38\pm0.06$ & $<0.12$       & \nodata       & $0.41\pm0.03$ \\
                &  $-200$ & $0.74\pm0.07$ & $0.49\pm0.05$ & \nodata       & $0.87\pm0.05$ \\
                &         & $0.98\pm0.12$ &               &               &               \\
{PG~$1407+265$} & $+2510$ & $<0.13$       & $<0.11$       & $0.21\pm0.03$ & $0.63\pm0.05$ \\
{PG~$1411+422$} &   $+80$ & $0.42\pm0.06$ & $0.38\pm0.09$ & \nodata       & $<0.78$\tablenotemark{f} \\
{PG~$1538+477$} &   $+90$ & $0.46\pm0.03$ & $0.35\pm0.02$ & $0.38\pm0.03$ & $0.62\pm0.03$ \\
{PG~$1700+518$} &   $+20$ & $1.09\pm0.14$ & $0.21\pm0.07$ & $<0.93$       & $1.09\pm0.08$ \\
{3C~$351.0$}\tablenotemark{d}
                & $-1290$ & $1.96\pm0.12$ & $0.76\pm0.36$ & $1.98\pm0.07$ & $2.23\pm0.29$ \\
                &  $-880$ & $0.10\pm0.06$ & $0.37\pm0.41$ & $0.25\pm0.05$ & $0.16\pm0.09$ \\
{H~$1821+643$}  &   $-20$ & $0.41\pm0.02$ & \nodata       & $0.15\pm0.04$ & $0.53\pm0.02$ \\
{PKS~$2251+11$} &   $610$ & $0.70\pm0.02$ & $0.46\pm0.01$ & $0.66\pm0.07$ & $0.74\pm0.00$ \\
{PG~$2302+029$} & $-4590$ & $0.20\pm0.03$ & $<0.04$       & $0.12\pm0.04$ & $0.56\pm0.03$ \\
\enddata
\tablenotetext{a}{We assume the convention of positive velocities for
absorption lines redshifted relative to the emission redshift. We
remind the reader that broad emission lines can be {\it blueshifted}
relative to the systemic redshift by over {$\sim1200$~\kms} (e.g.,
\cite{esp93}). Thus, positive velocities need not imply infall toward
the central engine.}
\tablenotetext{b}{All values are in the QSO rest--frame. The equivalent widths
of doublets are only those of the stronger transition
({C {\sc iv}$\lambda1548$}, {N {\sc v}$\lambda1238$}, and
{O {\sc vi}$\lambda1031$}). Equivalent width limits are {$3\sigma$}.}
\tablenotetext{c}{The two associated systems toward this QSO are apparently
line--locked in {C {\sc iv}}. We can, therefore, only constrain the
equivalent width of {C {\sc iv}$\lambda1548$} for the {$\Delta v_{\rm
abs}=-200$~\kms} system to a range defined by {$W_{\rm r}$} of the {C
{\sc iv}$\lambda1550$} line (first number) and {$W_{\rm r}$} of the
locked line (second number).}
\tablenotetext{d}{The associated system toward this QSO has two components
that are resolved. We list the deblended equivalent width of each
component. We do not treat the two components as separate systems in
the analysis.}
\tablenotetext{e}{Although we list only a limit for the {N {\sc v}} equivalent
width, we note that an absorption line does appear at the expected
wavelength.  However, we attribute this absorption to Galactic
{Si {\sc ii}$\lambda1527$}.}
\tablenotetext{f}{The {\Lya} line from the NAL is blended with that of the
mini--BAL. We list the total {\Lya} equivalent width as an upper
limit.}
\label{tab:nals}
\end{deluxetable}

\clearpage

\begin{deluxetable}{lcccrrrrrrccccl}
\rotate
\tablewidth{0pc}
\tabletypesize{\tiny}
\tablenum{3}
\tablecaption{QSO Properties}
\tablehead
{
\colhead{} &
\multicolumn{3}{c}{2500~\AA} &
\multicolumn{3}{c}{5~GHz} &
\multicolumn{2}{c}{2~keV} &
\colhead{} &
\colhead{} &
\colhead{} &
\colhead{} \\
\colhead{} &
\multicolumn{3}{c}{\hrulefill} &
\multicolumn{3}{c}{\hrulefill} &
\multicolumn{2}{c}{\hrulefill} &
\colhead{} &
\colhead{C {\sc iv}} &
\colhead{H$\beta$} &
\colhead{} &
\colhead{} &
\colhead{} \\
\colhead{Object} &
\colhead{$\log F_\nu$\tablenotemark{a}} &
\colhead{$\log L_\nu$\tablenotemark{a}} &
\colhead{$\alpha$} &
\colhead{$\log F_\nu$\tablenotemark{a}} &
\colhead{$\log L_\nu$\tablenotemark{a}} &
\colhead{$\alpha$} &
\colhead{$\log F_\nu$\tablenotemark{a}} &
\colhead{$\log L_\nu$\tablenotemark{a}} &
\colhead{$\alpha_{\mathrm{ox}}$} &
\colhead{\fwhm\tablenotemark{a}} &
\colhead{\fwhm\tablenotemark{a}} &
\colhead{$R_\mathrm{c}$} &
\colhead{Group} &
\colhead{Refs.\tablenotemark{b}}
}
\startdata
{PKS~$0003+15$}  & $-26.00$ & $30.67$ & $-0.67$ & $ -23.53$ & $ 33.14$ & $-0.76$ & $ -29.39$ & $ 27.27$ & $ -1.30$ & $ 2872$ & $ 4760$ & $ 0.32$ & 5 & 1, 2, 2, 3, 4 \\
{PG~$0043+039$}  & $-26.10$ & $30.42$ & $-2.53$ & $<-26.03$ & $<30.49$ & \nodata & $<-31.47$ & $<25.05$ & $<-2.06$ & \nodata & $ 5300$ & \nodata & 2 & 5, 6, 7, --, 8 \\
{PKS~$0044+030$} & $-26.36$ & $30.61$ & $-0.71$ & $ -24.26$ & $ 32.71$ & $-0.88$ & $ -30.10$ & $ 26.87$ & $ -1.43$ & $ 5253$ & $ 7700$ & $ 0.55$ & 5 & 9, 2, 10, 6, 4 \\
{PKS~$0122-00$}  & $-26.26$ & $31.23$ & $-1.36$ & $ -23.18$ & $ 34.31$ & $-0.25$ & $ -29.94$ & $ 27.54$ & $ -1.42$ & $ 7310$ & \nodata & $ 1.00$ & 4 & --, 2, 2, --, 4 \\
{3C~$57$}        & $-26.08$ & $30.96$ & $-0.51$ & $ -22.98$ & $ 34.06$ & $-0.66$ & $ -29.71$ & $ 27.33$ & $ -1.39$ & $ 5944$ & $ 4500$ & $ 0.73$ & 3 & 11, 2, 2, 3, 4 \\
{3C~$95$}        & $-26.06$ & $30.90$ & $-0.37$ & $ -23.07$ & $ 33.89$ & $-0.98$ & $ -29.83$ & $ 27.13$ & $ -1.45$ & $ 6557$ & $10000$ & $ 0.02$ & 5 & 12, 2, 2, 3, 4 \\
{PKS~$0405-12$}  & $-25.72$ & $31.18$ & $-0.54$ & $ -22.86$ & $ 34.03$ & $-0.41$ & $ -29.32$ & $ 27.58$ & $ -1.38$ & $ 2895$ & $ 3590$ & $ 0.32$ & 5 & 9, 2, 2, 3, 4 \\
{3C~$110$}       & $-25.97$ & $31.21$ & $-1.54$ & $ -23.56$ & $ 33.61$ & $-0.72$ & $ -29.72$ & $ 27.45$ & $ -1.44$ & $ 5856$ & $14970$ & $ 0.94$ & 4 & 13, 2, 2, 7, 4 \\
{PKS~$0439-433$} & $-26.54$ & $30.38$ & $-0.26$ & $ -22.70$ & $ 34.22$ & $ -0.1$ & $ -30.03$ & $ 26.89$ & $ -1.34$ & $ 2422$ & \nodata & $ 1.00$ & 3 & --, 2, 2, --, 4 \\
{HS~$0624+6907$} & $-25.60$ & $30.89$ & $-1.54$ & $<-26.03$ & $<30.46$ & \nodata & $ -29.50$ & $ 26.98$ & $ -1.50$ & $ 3553$ & $ 3276$ & \nodata & 1 & 14, --, --, --, 4 \\
{PKS~$0637-75$}  & $-26.17$ & $30.85$ & $-1.42$ & $ -22.50$ & $ 34.52$ & $ 0.08$ & $ -29.37$ & $ 27.65$ & $ -1.23$ & $ 2395$ & $ 5300$ & $ 1.00$ & 3 & 12, 2, 2, --, 4 \\
{B2~$0742+31$}   & $-26.22$ & $30.47$ & $-1.49$ & $ -23.14$ & $ 33.56$ & $-0.29$ & $ -29.78$ & $ 26.91$ & $ -1.37$ & $ 4545$ & $ 7720$ & $ 0.56$ & 3 & 1, 15, 10, 3, 4 \\
{US~$1867$}      & $-26.46$ & $30.33$ & $-0.88$ & $<-26.40$ & $<30.39$ & \nodata & $<-30.10$ & $<26.69$ & $<-1.40$ & $ 5247$ & $ 3080$ & \nodata & 1 & 16, --, --, --, 4 \\
{NGC~$2841$~UB3} & $-26.09$ & $30.77$ & $-1.26$ & $<-26.37$ & $<30.49$ & \nodata & $ -30.98$ & $ 25.88$ & $ -1.88$ & $ 4339$ & $ 4540$ & \nodata & 1 & 16, --, --, --, 17 \\
{PG~$0953+415$}  & $-25.83$ & $30.26$ & $-0.12$ & $<-25.89$ & $<30.20$ & $>0.6 $ & $ -29.69$ & $ 26.40$ & $ -1.48$ & $ 3321$ & $ 3130$ & $ 1.00$ & 1 & 13, 6, 18, --, 4 \\
{3C~$232$}       & $-26.20$ & $30.62$ & $-1.29$ & $ -23.29$ & $ 33.54$ & $-0.59$ & $ -31.19$ & $ 25.63$ & $ -1.92$ & $ 8015$ & $ 1310$ & $ 1.00$ & 4 & 1, 15, 10, 19, 17 \\
{$0959+68$~W1}   & $-26.10$ & $31.07$ & $-1.76$ & $ -24.34$ & $ 32.84$ & $-0.37$ & $ -30.31$ & $ 26.87$ & $ -1.61$ & $ 7064$ & \nodata & $ 0.49$ & 4 & --, 15, 10, 19, 17 \\
{TON~$28$}       & $-25.99$ & $30.39$ & $-0.35$ & $<-26.12$ & $<30.26$ & \nodata & $ -29.60$ & $ 26.77$ & $ -1.39$ & $ 3454$ & $ 2459$ & \nodata & 1 & 20, --, --, --, 4 \\
{4C~$41.21$}     & $-26.24$ & $30.71$ & $-0.19$ & $ -23.27$ & $ 33.69$ & $ 0.76$ & $ -29.65$ & $ 27.31$ & $ -1.31$ & $ 3743$ & $ 2600$ & $ 1.00$ & 3 & 12, 15, 21, 3, 4 \\
{PG~$1049-005$}  & $-26.16$ & $30.29$ & $-1.08$ & $<-26.03$ & $<30.43$ & \nodata & $ -30.09$ & $ 26.36$ & $ -1.51$ & \nodata & $ 5360$ & $ 0.52$ & 1 & 13, --, --, 6, 4 \\
{3C~$249.1$}     & $-26.06$ & $30.27$ & $-1.22$ & $ -23.09$ & $ 33.23$ & $-0.85$ & $ -29.33$ & $ 26.99$ & $ -1.26$ & $ 3175$ & $ 6600$ & $ 0.12$ & 5 & 12, 22, 22, 6, 4 \\
{MC~$1104+167$}  & $-26.09$ & $30.90$ & $-0.81$ & $ -23.35$ & $ 33.64$ & $-0.24$ & $ -29.52$ & $ 27.47$ & $ -1.32$ & $ 4250$ & $ 4610$ & $ 0.55$ & 3 & 13, 2, 10, 3, 4 \\
{PG~$1116+215$}  & $-25.41$ & $30.41$ & $-0.83$ & $ -25.48$ & $ 30.33$ & $-0.66$ & $ -29.30$ & $ 26.52$ & $ -1.49$ & $ 4669$ & $ 2920$ & $ 0.69$ & 1 & 9, 6, 7, 6, 4 \\
{$1130+106$~Y}   & $-26.45$ & $30.33$ & $-0.92$ & $<-26.05$ & $<30.73$ & \nodata & $ -30.13$ & $ 26.65$ & $ -1.41$ & $ 5907$ & \nodata & \nodata & 1 & --, --, --, --, 4 \\
{PKS~$1136-13$}  & $-26.35$ & $30.51$ & $-0.81$ & $ -22.72$ & $ 34.14$ & $-0.47$ & $ -29.88$ & $ 26.98$ & $ -1.36$ & $ 2797$ & $ 2670$ & $ 0.12$ & 5 & 13, 2, 2, 3, 4 \\
{3C~$263$}       & $-26.06$ & $30.95$ & $-0.88$ & $ -22.56$ & $ 34.45$ & $ 0.55$ & $ -29.74$ & $ 27.27$ & $ -1.41$ & $ 2766$ & $ 5560$ & $ 0.11$ & 3 & 13, 23, 10, 3, 4 \\
{PG~$1216+069$}  & $-25.99$ & $30.40$ & $-0.36$ & $<-25.61$ & $<30.78$ & $>0.54$ & $ -29.63$ & $ 26.76$ & $ -1.40$ & $ 3071$ & $ 5190$ & $ 1.00$ & 1 & 13, 6, 7, 6, 4 \\
{Mrk~$205$}      & $-25.59$ & $29.41$ & $-0.87$ & $ -25.22$ & $ 29.78$ & \nodata & $ -29.19$ & $ 25.81$ & $ -1.38$ & $ 3305$ & $ 3330$ & \nodata & 2 & 9, --, --, --, 4 \\
{3C~$273$}       & $-24.63$ & $31.08$ & $-0.87$ & $ -21.37$ & $ 34.34$ & $-0.11$ & $ -28.36$ & $ 27.35$ & $ -1.43$ & $ 3599$ & $ 3520$ & $ 0.71$ & 3 & 9, 2, 2, 6, 4 \\
{PG~$1248+401$}  & $-26.46$ & $30.99$ & $-0.18$ & $<-26.43$ & $<31.02$ & \nodata & $ -30.15$ & $ 27.30$ & $ -1.42$ & $ 7897$ & \nodata & \nodata & 4 & --, --, --, 6, 4 \\
{PKS~$1252+11$}  & $-26.42$ & $30.87$ & $-1.14$ & $ -23.38$ & $ 33.91$ & $-0.02$ & $ -30.13$ & $ 27.16$ & $ -1.43$ & $ 4559$ & \nodata & $ 0.89$ & 3 & --, 2, 2, 19, 4 \\
{PG~$1254+047$}  & $-26.28$ & $31.16$ & $-0.77$ & $<-26.11$ & $<31.33$ & \nodata & $<-30.78$ & $<26.66$ & $<-1.73$ & $ 6008$ & \nodata & \nodata & 4 & --, --, --, --, 24 \\
{PG~$1259+593$}  & $-25.90$ & $30.81$ & $-0.93$ & $<-26.04$ & $<30.67$ & \nodata & $<-30.33$ & $<26.38$ & $<-1.70$ & $ 5628$ & $ 3390$ & \nodata & 1 & 5, --, --, --, 25 \\
{PG~$1302-102$}  & $-25.61$ & $30.64$ & $-1.13$ & $ -23.15$ & $ 33.10$ & $ 0.27$ & $ -29.60$ & $ 26.65$ & $ -1.53$ & $ 2826$ & $ 3400$ & $ 1.00$ & 3 & 13, 2, 7, --, 4 \\
{TON~$153$}      & $-26.10$ & $31.34$ & $-0.58$ & $ -25.79$ & $ 31.65$ & \nodata & $ -30.56$ & $ 26.88$ & $ -1.71$ & $ 5479$ & \nodata & \nodata & 4 & --, --, --, --, 26 \\
{PG~$1333+176$}  & $-26.07$ & $30.79$ & $-1.36$ & $ -24.71$ & $ 32.15$ & $-0.37$ & $ -30.06$ & $ 26.80$ & $ -1.53$ & $ 6346$ & $ 3680$ & $ 1.00$ & 4 & 16, 6, 7, 6, 24 \\
{PKS~$1354+19$}  & $-26.21$ & $30.90$ & $-1.30$ & $ -22.92$ & $ 34.19$ & $-0.34$ & $ -30.00$ & $ 27.11$ & $ -1.46$ & $ 2244$ & $ 4710$ & $ 1.00$ & 3 & 13, 2, 2, 3, 4 \\
{PG~$1407+265$}  & $-26.16$ & $31.21$ & $-0.44$ & $ -25.33$ & $ 32.03$ & $-0.11$ & $ -29.88$ & $ 27.48$ & $ -1.43$ & $ 6595$ & \nodata & $ 0.37$ & 4 & --, 6, 7, 6, 4 \\
{PG~$1411+442$}  & $-25.75$ & $29.45$ & $-1.25$ & $ -26.18$ & $ 29.02$ & $-1.09$ & $ -31.06$ & $ 24.14$ & $ -2.04$ & $ 2037$ & $ 2670$ & $ 0.85$ & 2 & 13, 6, 7, 6, 17 \\
{PKS~$1424-11$}  & $-26.32$ & $30.89$ & $-1.91$ & $ -23.41$ & $ 33.80$ & $-1.14$ & $ -30.08$ & $ 27.14$ & $ -1.44$ & $ 3840$ & \nodata & $ 0.00$ & 5 & --, 2, 2, --, 4 \\
{PG~$1444+407$}  & $-25.83$ & $30.36$ & $-0.58$ & $<-26.34$ & $<29.85$ & \nodata & $ -30.08$ & $ 26.10$ & $ -1.63$ & $ 4060$ & $ 2480$ & \nodata & 1 & 13, --, --, --, 4 \\
{B2~$1512+37$}   & $-26.27$ & $30.21$ & $-0.14$ & $ -23.47$ & $ 33.02$ & $-0.71$ & $ -29.70$ & $ 26.79$ & $ -1.32$ & $ 3947$ & $ 6810$ & $ 0.16$ & 5 & 13, 15, 10, 3, 4 \\
{PG~$1538+477$}  & $-25.77$ & $31.40$ & $-2.07$ & $ -24.11$ & $ 33.07$ & $ 0.61$ & $ -30.07$ & $ 27.10$ & $ -1.65$ & $ 5587$ & $ 4920$ & $ 0.73$ & 4 & 16, 6, 10, 6, 4 \\
{3C~$334.0$}     & $-26.22$ & $30.64$ & $-0.96$ & $ -23.22$ & $ 33.64$ & $  -1$  & $ -30.01$ & $ 26.85$ & $ -1.45$ & $ 6045$ & $ 9400$ & $ 0.19$ & 5 & 13, 2, 2, 3, 4 \\
{PG~$1700+518$}  & $-25.92$ & $30.35$ & $-2.29$ & $ -25.13$ & $ 31.13$ & $-0.88$ & $<-31.68$ & $<24.58$ & $<-2.21$ & \nodata & $ 2210$ & $ 0.29$ & 2 & 13, 6, 7, 6, 27 \\
{3C~$351.0$}     & $-25.94$ & $30.55$ & $-0.84$ & $ -22.94$ & $ 33.55$ & $-0.78$ & $ -30.25$ & $ 26.24$ & $ -1.65$ & $ 4355$ & $ 6560$ & $ 0.01$ & 5 & 9, 23, 10, 3, 17 \\
{H~$1821+643$}   & $-25.34$ & $30.94$ & $-1.11$ & $<-26.02$ & $<30.27$ & \nodata & $ -28.87$ & $ 27.41$ & $ -1.36$ & $ 3640$ & $ 6120$ & \nodata & 1 & 9, --, --, --, 4 \\
{4C~$73.18$}     & $-26.12$ & $30.18$ & $-1.68$ & $ -22.52$ & $ 33.78$ & $0.004$ & $ -29.41$ & $ 26.89$ & $ -1.26$ & $ 3303$ & $ 3360$ & $ 1.00$ & 3 & 13, 15, 10, --, 4 \\
{PG~$2112+059$}  & $-26.18$ & $30.51$ & $-1.44$ & $ -26.02$ & $ 30.66$ & $-1.05$ & $ -31.15$ & $ 25.53$ & $ -1.91$ & $ 4863$ & $ 3190$ & $ 0.00$ & 2 & 13, 6, 7, 6, 17 \\
{PKS~$2128-12$}  & $-26.01$ & $30.76$ & $-1.06$ & $ -22.90$ & $ 33.87$ & $ 0.08$ & $ -29.48$ & $ 27.29$ & $ -1.33$ & $ 4356$ & $ 7050$ & $ 0.95$ & 3 & 1, 2, 2, 29, 4 \\
{PKS~$2145+06$}  & $-26.22$ & $31.19$ & $-1.27$ & $ -22.81$ & $ 34.60$ & $ 0.32$ & $ -29.74$ & $ 27.67$ & $ -1.35$ & $ 6655$ & \nodata & $ 0.64$ & 4 & --, 2, 2, 19, 4 \\
{PKS~$2243-123$} & $-26.31$ & $30.67$ & $-0.71$ & $ -23.01$ & $ 33.97$ & $-0.23$ & $<-29.87$ & $<27.11$ & $<-1.37$ & $ 2495$ & $ 2118$ & $ 0.95$ & 3 & 28, 2, 2, 29, 4 \\
{3C~$454.3$}     & $-26.01$ & $31.26$ & $-2.19$ & $ -22.04$ & $ 35.24$ & $ 0.51$ & $ -28.96$ & $ 28.32$ & $ -1.13$ & $ 2878$ & \nodata & $ 0.98$ & 4 & --, 2, 2, 3, 4 \\
{PKS~$2251+11$}  & $-26.16$ & $30.20$ & $-0.80$ & $ -23.34$ & $ 33.02$ & $-0.93$ & $ -30.47$ & $ 25.89$ & $ -1.65$ & $ 4138$ & $ 3170$ & $ 0.00$ & 2 & 12, 2, 2, 6, 24 \\
{PKS~$2300-683$} & $-26.66$ & $30.13$ & $-0.67$ & $ -22.57$ & $ 34.22$ & $-0.45$ & $ -30.11$ & $ 26.68$ & $ -1.33$ & $ 2937$ & $ 2230$ & $ 1.00$ & 3 & 1, 2, 2, --, 4 \\
{PG~$2302+029$}  & $-25.82$ & $31.65$ & $-1.73$ & $<-26.09$ & $<31.38$ & $-0.38$ & $<-29.83$ & $<27.64$ & $<-1.54$ & $11180$ & \nodata & $ 1.00$ & 4 & --, 30, 30, --, 4 \\
{PKS~$2340-036$} & $-26.16$ & $31.16$ & $-0.77$ & $ -23.77$ & $ 33.54$ & $-0.22$ & $ -29.58$ & $ 27.74$ & $ -1.31$ & $ 4922$ & \nodata & $ 1.00$ & 3 & --, 2, 10, --, 4 \\
{PKS~$2344+09$}  & $-26.21$ & $30.84$ & $-1.59$ & $ -23.06$ & $ 33.99$ & $-0.31$ & $ -29.99$ & $ 27.06$ & $ -1.45$ & $ 2664$ & $ 2360$ & $ 1.00$ & 3 & 13, 2, 2, 6, 4 \\
{PKS~$2352-342$} & $-26.30$ & $30.78$ & $-0.92$ & $ -23.66$ & $ 33.42$ & $ 0.17$ & $ -29.81$ & $ 27.27$ & $ -1.35$ & $ 3191$ & \nodata & $ 1.00$ & 3 & --, 2, 2, --, 4 \\
\enddata
\tablenotetext{a}{$\rm [F_\nu]=erg~cm^{-2}~s^{-1}~Hz^{-1}$,~
$\rm [L_\nu]=erg~s^{-1}~Hz^{-1}$,~[{\fwhm}]=\kms}
\tablenotetext{b}{
The code of numbers refers to the references for the {\Hb~\fwhm} (first
number), the radio spectral index (second and third numbers), the
radio core fraction (fourth number), and the soft X--ray count rate (fifth
number). A dash indicates that the information was not available at the time
of compilation and, thus, no reference is given.}
\tablerefs{(1) Corbin (1997\nocite{cor97}); (2) Wright \& Otrupeck
(1990\nocite{pkscat}); (3) Nilsson (1998\nocite{nil98}); (4) Voges
{\etal} (1999\nocite{rassbsc}); (5) Boroson \& Green
(1992\nocite{bg92}); (6) Kellermann {\etal} (1989\nocite{kel89}); (7)
Condon {\etal} (1998\nocite{nvss}); (8) Gallagher {\etal}
(1999\nocite{gallsc}); (9) Corbin (1991\nocite{cor91}); (10) White \&
Becker (1992\nocite{wb92}); (11) Brotherton (1996\nocite{bro96}); (12)
Wills \& Browne (1986\nocite{wb86}); (13) Corbin \& Boroson
(1996\nocite{cb96}); (14) Reimers {\etal} (1995\nocite{rei95}); (15)
Gregory \& Condon (1991\nocite{gs91}); (16) Marziani {\etal}
(1996\nocite{mar96}); (17) White {\etal} (1994\nocite{wgacat}); (18)
White {\etal} (2000\nocite{second}); (19) Laurent--Muehleisen {\etal}
(1997\nocite{sallylm97}); (20) Miller {\etal} (1992\nocite{mil92});
(21) Becker {\etal} (1995\nocite{first}); (22) Kellermann {\etal}
(1969\nocite{kpw69}); (23) K\"uhr {\etal} (1981\nocite{kuhr81}); (24)
Tananbaum {\etal} (1986\nocite{tea}); (25) Brandt {\etal}
(2000\nocite{brandt}); (26) Voges {\etal} (1995\nocite{rosatsrc});
(27) Green \& Mathur (1996\nocite{gm96}); (28) Tadhunter {\etal}
(1993\nocite{tad93}); (29) Morganti {\etal} (1997\nocite{mor97}); (30)
Peacock {\etal} (1986\nocite{pml86})}
\label{tab:props}
\end{deluxetable}

\clearpage

\begin{deluxetable}{rcccccccc}
\tabletypesize{\footnotesize}
\tablewidth{0pc}
\tablenum{4}
\tablecaption{Spearmann Correlation Probabilities}
\tablehead
{
\colhead{} &
\multicolumn{2}{c}{$\log L_\nu$} &
\multicolumn{2}{c}{$\alpha$} &
\colhead{} &
\colhead{} &
\colhead{} &
\colhead{} \\
\colhead{} &
\multicolumn{2}{c}{\hrulefill} &
\multicolumn{2}{c}{\hrulefill} &
\colhead{$\rm \log~H\beta$} &
\colhead{$\log~$C {\sc iv}} &
\colhead{} &
\colhead{} \\
\colhead{} &
\colhead{5~GHz} &
\colhead{2~keV} &
\colhead{2500~\AA} &
\colhead{5~GHz} &
\colhead{\fwhm} &
\colhead{\fwhm} &
\colhead{$R_{\rm c}$} &
\colhead{$W_{\mathrm{r}}($C {\sc iv}$)$}
}
\startdata
{$\log L_\nu($2500~\AA$)$}  & $0.005$ & $3\times10^{-8}$ & $0.175$ & $0.026$ & $0.030$ & $0.005$ & $0.474$ & $0.794$ \\
{$\log L_\nu(\mathrm{5~GHz})$}     & \nodata & $2\times10^{-6}$ & $0.531$ & $0.004$ & $0.155$ & $0.103$ & $10^{-4}$ & $0.602$ \\
{$\log L_\nu(\mathrm{2~keV})$}     & \nodata & \nodata & $0.927$ & $6\times10^{-4}$ & $0.012$ & $0.884$ & $0.011$ & $0.019$ \\
{$\alpha($2500~\AA$)$}      & \nodata & \nodata & \nodata & $0.961$ & $0.520$ & $0.492$ & $0.393$ & $0.273$ \\
{$\alpha(\mathrm{5~GHz})$}         & \nodata & \nodata & \nodata & \nodata &$0.810$ & $0.073$ & $2\times10^{-5}$ & $0.112$ \\
{$\rm \log$~H$\beta$~\fwhm}         &  \nodata & \nodata & \nodata & \nodata & \nodata & $0.137$ & $0.627$ & $0.437$ \\
{$\log~$C {\sc iv}~\fwhm}           &  \nodata & \nodata & \nodata & \nodata & \nodata & \nodata & $0.121$ & $0.789$ \\
{$R_{\mathrm{c}}$}                       &  \nodata & \nodata & \nodata & \nodata & \nodata & \nodata & \nodata & $0.087$ \\
\enddata
\label{tab:spear}
\end{deluxetable}

\clearpage

\begin{deluxetable}{lrrrr}
\tablewidth{0pc}
\tablenum{5}
\tablecaption{Kolmogorov--Smirnov Logarithmic Probabilities}
\tablehead
{
\colhead{} &
\colhead{1,2 vs.} &
\colhead{3,4} &
\colhead{1 vs.} &
\colhead{3 vs.} \\
\colhead{Property} &
\colhead{3,4,5} &
\colhead{vs. 5} &
\colhead{2} &
\colhead{4}
}
\startdata
$\log L_\nu($2500~\AA$)$ & $-3.7$ & $-1.4$ & $-0.6$ & $-3.2$ \\
$\log L_\nu($5~GHz$)$    & $-8.4$ & $-0.3$ & $-0.3$ & $-2.2$ \\
$\log L_\nu($2~keV$)$    & $-6.1$ & $-0.8$ & $-2.6$ & $-0.3$ \\
$\rm \alpha($2500~\AA$)$       & $ 0.0$ & $-0.5$ & $-0.6$ & $-0.8$ \\
$\rm \alpha($5~GHz$)$          & $-3.0$ & $-3.6$ & $-2.6$ & $-1.4$ \\
{\Hb~\fwhm}                  & $-2.4$ & $-1.4$ & $-0.1$ & $-0.5$ \\
{\CIV~\fwhm}                 & $-0.6$ & $-0.1$ & $ 0.0$ & $-4.6$ \\
$R_{\rm c}$                  & $-3.2$ & $-4.0$ & $-0.1$ & $-1.7$ \\
\enddata
\label{tab:clusprops}
\end{deluxetable}

\clearpage

\begin{deluxetable}{cccccccl}
\rotate
\tablewidth{0pc}
\tabletypesize{\footnotesize}
\tablenum{6}
\tablecaption{Properites of Clusters}
\tablehead
{
\colhead{} &
\colhead{} &
\colhead{} &
\colhead{} &
\colhead{Expected} &
\multicolumn{2}{c}{Probabilities} &
\colhead{} \\
\colhead{} &
\colhead{} &
\colhead{} &
\colhead{NALQSO} &
\colhead{Number} &
\multicolumn{2}{c}{\hrulefill} &
\colhead{} \\
\colhead{Cluster} &
\colhead{QSOs} &
\colhead{NALQSOs} &
\colhead{fraction\tablenotemark{a}} &
\colhead{of NALs\tablenotemark{b}} &
\colhead{Interv.} &
\colhead{Assoc.} &
\colhead{Main Distinguishing Properties}
} 
\startdata
1 & 12 & 4 & $33\pm14$\% & 1.0 & $0.0110$ & $0.1978$ & radio--quiet,
flat radio spectra\\
2 & 6 & 4 & $67\pm19$\% & 0.5 & $0.0006$ & $0.0348$ & radio--quiet,
steep radio spectra, UV and X--ray weak\\
3 & 18 & 0 & $0$\% & 1.5 & $0.2171$ & $0.0051$ & radio--loud, flat
radio spectra\\
4 & 13 & 5 & $38\pm13$\% & 1.0 & $0.0023$ & $0.1307$ & radio--loud,
flat radio spectra, broad {\CIV} \\
5 & 10 & 2 & $20\pm13$\% & 0.8 & $0.1511$ & $0.2783$ & radio--loud,
steep radio spectra\\ \hline Total & 59 & 15 & $25\pm6$\% & 5.0 &
$4.3\times10^{-5}$ & $0.1186$ & \\
\enddata
\tablenotetext{a}{The formal {1$\sigma$} errors were computed assuming
a binomial distribution of QSOs.}
\tablenotetext{b}{In principle, it would be better to know the number of
{\it intervening} NALs expected in each group, this is not possible given
the Jannuzi {\etal} (1998) line list, since it is not known how many are
intrinsic to the QSO. (Such knowledge, of course, would subvert the need
for the computation.)}
\label{tab:cluster}
\end{deluxetable}

\clearpage

\end{document}